\documentclass[aps,pre,reprint,showpacs,superscriptaddress]{revtex4-2}
\usepackage[colorlinks,linkcolor=blue,anchorcolor=blue,citecolor=blue,urlcolor=blue]{hyperref}
\usepackage{amsmath}
\usepackage{graphicx}
\usepackage{dcolumn}
\usepackage{float}
\usepackage{enumitem}
\begin{document}
	\title{Coexistence of positive and negative information in information-epidemic dynamics on multiplex networks}
	\author{Li-Ying Liu}
	\address{School of Physics, Northwest University, Xi'an 710127, China}
	\author{Chao-Ran Cai}\email{ccr@nwu.edu.cn}
	\address{School of Physics, Northwest University, Xi'an 710127, China}
	\address{Shaanxi Key Laboratory for Theoretical Physics Frontiers, Xi'an 710127, China}
	\author{Si-Ping Zhang}
	\address{Key Laboratory of Biomedical Information Engineering of Ministry of Education, and Institute of Health and Rehabilitation Science, School of Life Science and Technology, Xi’an Jiaotong University, Xi’an 710049, China}
	\author{Bin-Quan Li}
	\address{School of Physics, Ningxia University, Yinchuan 750021, China}
	
	\date{\today}
	\begin{abstract}
This paper investigates the coexistence of positive and negative information in the context of information-epidemic dynamics on multiplex networks.
In accordance with the tenets of mean field theory, we present not only the analytic solution of the prevalence threshold, but also the coexistence conditions of two distinct forms of information (i.e., the two phase transition points at which a single form of information becomes extinct).
In regions where multiple forms of information coexist, two completely distinct patterns emerge: monotonic and non-monotonic. 
The physical mechanisms that give rise to these different patterns have also been elucidated.
The theoretical results are robust with regard to the network structure and show a high degree of agreement with the findings of the Monte Carlo simulation.
	\end{abstract}
	\maketitle
	\section{INTRODUCTION}
In its most general sense, information refers to anything that is transmitted by human community, including opinions, knowledge, ideas, thoughts, news, and so on.	
Diversity is one of the most obvious characteristics of information.
There are numerous reasons for this diversity~\cite{doi:10.1073/pnas.1805871115}, such as the varying cognitive depth of things, the disparate perspectives on problems, the individual differences in understanding abilities, and the direct falsification of information content.
Recently, many researchers have devoted their efforts to the investigation of the coexistence and evolution of multiple information on networks~\cite{PhysRevE.81.056102,Wang_2012,PhysRevE.93.032305,WANG201763,doi:10.1126/science.aap9559,doi:10.1126/science.aao2998,PhysRevLett.124.048301,YAO2020122764,ZHANG2022366,PhysRevE.105.024125,PhysRevE.106.014301}.
Trpevski et al. found that information of different priorities may coexist within a system as a consequence of network attributes, such as a low average degree and a small clustering coefficient~\cite{PhysRevE.81.056102}.
Wang et al. examined the exclusive and non-exclusive influences of the two informations and discovered that the non-exclusive characteristic is conducive to coexistence~\cite{Wang_2012}.
As research progresses, targeted researches have been conducted on false or negative information in diversity.
Vosoughi's empirical research, based on data from Twitter, revealed that false news is more prevalent online than the truth~\cite{doi:10.1126/science.aap9559}.
Meanwhile, many strategies have been proposed to control the dissemination of negative information on online social networks~\cite{8717734,9036077,10336945,Chu2024}.
Some researchers studied the evolution of multiple information from a game theory perspective~\cite{PhysRevE.106.034303,9915429}.

It is also important to consider the feedback loop between the information diffusion and human behavior.
For example, the impact of negative word-of-mouth on the net present value of the firm has been demonstrated to be significant~\cite{GOLDENBERG2007186}. 
The spread of misinformation leads to poor utilization of vaccines~\cite{loomba2021measuring,doi:10.1126/science.adk3451}, while accurate information about descriptive norms can increase intentions to accept vaccines~\cite{Moehring2023}.
It is well known that the relationship between disease dynamics and human behavior is complex~\cite{Ferguson2007,WANG20151}, and the disease-related information plays a role in mediating this relationship.
The study of this feedback effect is exemplified by the coupled information-epidemic dynamics on multiplex networks~\cite{PhysRevLett.111.128701,PhysRevE.90.012808,PhysRevE.104.044303,PhysRevResearch.5.013196}.
The feedback mechanism of information-epidemic dynamics is as follows: individuals will obtain pertinent information about disease through a variety of sources and will implement efficacious protective measures to mitigate the risk of infection; those infected with the virus confirm the veracity of the information and facilitate its dissemination due to their empathy.
In light of this feedback mechanism, researchers have undertaken a series of investigations into the impact of homogeneous information on disease dynamics from a range of perspectives in the past decade~\cite{Zhou_2019,PhysRevE.100.032313,PhysRevE.102.022312,kabir2020impact,PhysRevE.106.034307,PhysRevE.102.042314,PhysRevResearch.3.013157,GUO2021127282,guo2021impact,10.1063/5.0099183,LIU2023113657,Yin2023,PhysRevResearch.5.033220,Masoomy_2023,PhysRevResearch.5.033065,CHANG2024114780}, including theoretical analysis method~\cite{PhysRevLett.111.128701,Zhou_2019,PhysRevE.104.044303}, mass media~\cite{PhysRevE.90.012808,PhysRevE.106.034307}, vaccination~\cite{kabir2020impact}, network topology~\cite{guo2021impact,PhysRevE.104.044303,GUO2021127282}, relative timescale~\cite{PhysRevE.100.032313,PhysRevE.102.022312,PhysRevResearch.5.033220,CHANG2024114780}, and higher-order interactions~\cite{PhysRevResearch.5.013196,LIU2023113657,10.1063/5.0099183}.

As previously stated, human communication often results in the generation of a considerable amount of information noise, misinformation, and even false information~\cite{doi:10.1126/science.aao2998,gallotti2020assessing}.
Meanwhile, the risk of infection varies from one individual to another, depending on the specific details of the information and the subsequent behavioral response.
For instance, the risk of infection differs significantly between individuals who do not use protection, those who wear masks, and those who wear masks and avoid crowded places.
More recently, the various research interests that have been opened up by multiple information dynamics and information-epidemic dynamics are beginning to integrate~\cite{8957067,HUANG2021125536}.
The current state of research in this field is still in its early stages, with the majority of studies concentrating on the influence of information dissemination on the spread of epidemics~\cite{Chen2022CoevolvingSD,10.1063/5.0126799,FANG2023113376,LI2024128700,HAN2024115264}. 
For instance, the promotion of positive information (which is more effective in reducing the risk of infection) has the potential to contribute to the control of epidemic outbreaks.
Nevertheless, the coexistence of multiple information represents a significant challenge, as it provides insight into the circumstances under which the model may revert to a classical information-epidemic model.

In this paper, we discuss the coexistence of two kinds of information based on the information-epidemic dynamics of multiplex networks.
In the case of a system situated far from the epidemic threshold, with an increase in the infection rate, the epidemic prevalence exhibit two completely distinct patterns (monotony and non-monotony) in the area of the coexistence of multiple information.
Based on the mean field theory, we give the conditions for the coexistence of multiple information, which is achieved by determining the phase transition point of the respective information annihilation.
Furthermore, we elucidate the physical mechanisms underlying the disparate patterns observed in epidemic prevalence.

The rest of this paper is organized as follows. 
In Sec.~\ref{model}, we provide a comprehensive account of the $\mathrm{UA_1A_2U}$-SIS model, which comprises two distinct categories of information.
In Sec.~\ref{lilun}, we present the dynamic equations of the model and derive the analytic solution of the epidemic threshold.
In Sec.~\ref{result}, we discuss the coexistence of two kinds of information near and far from the epidemic threshold, and give their coexistence conditions.
In Sec.~\ref{simulation}, we compare the theoretical results with Monte Carlo simulations.
In Sec.~\ref{conclusion}, we summarize our results.

\section{Model}\label{model}
\begin{figure*}
	\includegraphics[width=\linewidth]{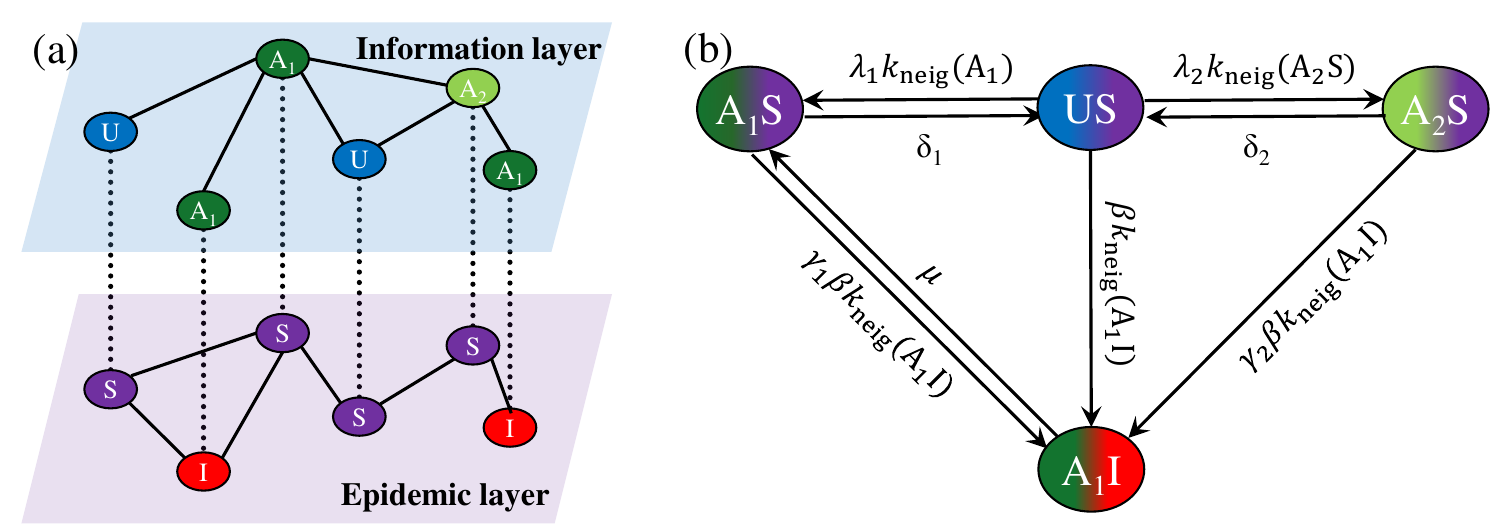}
	\caption{The schematic illustrations demonstrate the $\mathrm{UA_1A_2U}$-SIS model on a multiplex network in (a), along with the associated transition rates between states in (b).
The information layer describes the diffusion of information that facilitates the adoption of protective measures against the disease, and nodes have three kinds of states: unaware (U), aware of positive preventive information ($\mathrm{A_1}$), and aware of negative preventive information ($\mathrm{A_2}$).
The epidemic layer describes the spread of disease in populations with heterogeneous protective measures, and nodes can be either infected (I) or susceptible (S).
It should be noted that UI and $\mathrm{A_2I}$ are excluded from the $\mathrm{UA_1A_2U}$-SIS model. Consequently, individuals can be in four possible states: $\mathrm{A_1S}$, $\mathrm{A_2S}$, US, and $\mathrm{A_1I}$.
The quantity $k_{\mathrm{neig}}(\mathrm{X})$ represents the number of the individual's neighbors who are in the state $X$, where $\mathrm{X\in\{A_1S, A_2S, US, A_1I}\}$ and $k_{\mathrm{neig}}(\mathrm{A_1})=k_{\mathrm{neig}}(\mathrm{A_1S})+k_{\mathrm{neig}}(\mathrm{A_1I})$.}	\label{fig:fig1}
\end{figure*}

Two-layered multiplex networks are employed to the $\mathrm{UA_1A_2U}$-SIS model~\cite{8957067}, comprising an epidemic layer and an information layer. 
As shown in Fig.~\ref{fig:fig1}(a), nodes in different layers represent the same individuals, but they may be connected differently in the two layers. 
The epidemic layer represents the propagation of epidemics in physical contact networks, while the information layer describes the diffusion of information related to the infectious diseases via social networks or physical and social hybrid networks.

Individuals on the epidemic layer are classified into two types of compartments: susceptible (S) and infected (I). 
Nodes on the information layer are divided into three kinds of compartments: unaware (U), aware of positive preventive information ($\mathrm{A_1}$), and aware of negative preventive information ($\mathrm{A_2}$).
Aware individuals (i.e., $\mathrm{A_1}$ and $\mathrm{A_2}$) take protective measures to avoid infection, but $\mathrm{A_1}$ takes stronger measures than $\mathrm{A_2}$.
The following three assumptions are employed in this paper. First, it is assumed that individuals are aware of the infection when they become infected~\cite{PhysRevLett.111.128701,10.1063/5.0126799}. Second, it is assumed that infected individuals often disseminate positive information to help others avoid further infection~\cite{8957067,10.1063/5.0126799,LI2024128700}. Third, it is also
assumed that once an individual accepts a certain kind of preventive information, this individual will not accept any other information~\cite{PhysRevE.81.056102,8957067}.
Therefore, individuals in the multiplex network can be divided into four different classes: susceptible and aware of positive information ($\mathrm{A_1S}$), susceptible and aware of negative information ($\mathrm{A_2S}$), susceptible and unaware (US), and infected and aware of positive information ($\mathrm{A_1I}$).

The diffusion of positive information occurs at a rate of $\lambda_1$ in each edge US-$\mathrm{A_1S}$ and US-$\mathrm{A_1I}$, while the dissemination of negative information occurs at a rate of $\lambda_2$ in each edge US-$\mathrm{A_2S}$.
Those with states $\mathrm{A_1S}$ and $\mathrm{A_2S}$ have forgotten the existence of infectious disease and become the unaware individuals at rates of $\delta_1$ and $\delta_2$, respectively.
The heterogeneity of information possessed by individuals results in a transmission rate of the disease on each edge of US-$\mathrm{A_1I}$, $\mathrm{A_1S}$-$\mathrm{A_1I}$, and $\mathrm{A_2S}$-$\mathrm{A_1I}$ being $\beta$, $\gamma_1\beta$, and $\gamma_2\beta$, respectively.
Here, the qualitative relationship between the two attenuation coefficients is defined as $0\le\gamma_1<\gamma_2\le1$, reflecting the observation that individuals in $\mathrm{A_1}$ adopt more robust protective measures than those in $\mathrm{A_2}$.
In particular, when $\gamma_1$ is 0, the $\mathrm{A_1}$ individuals are fully immune to the infectious disease.
Finally, individuals infected with the virus ($\mathrm{A_1I}$) are cured to the $\mathrm{A_1S}$ class again at a rate of $\mu$. 
For the sake of clarity and convenience, all possible state transitions of individuals are shown in Fig.~\ref{fig:fig1}(b).

This paper adopts continuous-time dynamics, which can avoid the priority judgment problem in discrete-time dynamics because individuals cannot hold two types of information at the same time.


\section{Theoretical analysis }\label{lilun}

\subsection{Individual-based mean-field theory on multiplex networks}
The individual-based theory permits the precise calculation of the exact expected probabilities of each individual state by fully considering the adjacency matrix~\cite{pastor2015epidemic}.
The incorporation of the assumption that the states of adjacent nodes are statistically independent (which also called the absence of dynamical correlations~\cite{PhysRevLett.116.258301}) gives rise to the individual-based mean-field theory, also known as the quenched mean-field theory~\cite{PhysRevE.85.056111} and the microscopic Markov chain theory~\cite{PhysRevLett.111.128701}.

Let us denote $[a_{ji}]$ and $[b_{ji}]$ the adjacency matrices of the information and epidemic layer in the $\mathrm{UA_1A_2U}$-SIS model, respectively.
For each individual $i$, there is a certain probability of being in one of the four states at time $t$. These probabilities are denoted by $P_i^{\mathrm{US}}(t)$, $P_i^{\mathrm{A_1S}}(t)$, $P_i^{\mathrm{A_2S}}(t)$, and $P_i^{\mathrm{A_1I}}(t)$, respectively.
Using the scheme presented in Fig.~\ref{fig:fig1}(b), the dynamic equations for each individual $i$ can be written as
\begin{subequations}\label{eqq1}
\begin{align}\label{eqq1a}\nonumber
\frac{dP_i^{\mathrm{A_1S}}(t)}{dt}=&P_i^{\mathrm{A_1I}}(t)\mu+P_i^{\mathrm{US}}(t)\lambda_1\sum_{j}P_j^{\mathrm{A_1}}(t)a_{ji} \\ 
&-P_i^{\mathrm{A_1S}}(t)\left[\delta_1+\gamma_1\beta\sum_{j}P_j^{\mathrm{A_1I}}(t)b_{ji}\right],\\\label{eqq1b} \nonumber
\frac{dP_i^{\mathrm{A_2S}}(t)}{dt}=&P_i^{\mathrm{US}}(t)\lambda_2\sum_{j}P_j^{\mathrm{A_2S}}(t)a_{ji}\\ 
&-P_i^{\mathrm{A_2S}}(t)\left[\delta_2+\gamma_2\beta\sum_{j}P_j^{\mathrm{A_1I}}(t)b_{ji}\right],\\\label{eqq1c} \nonumber
\frac{dP_i^{\mathrm{A_1I}}(t)}{dt}=&-P_i^{\mathrm{A_1I}}(t)\mu+P_i^{\mathrm{US}}(t)\beta\sum_{j}P_j^{\mathrm{A_1I}}(t)b_{ji}\\ \nonumber
&+P_i^{\mathrm{A_1S}}(t)\gamma_1\beta\sum_{j}P_j^{\mathrm{A_1I}}(t)b_{ji}\\
&+P_i^{\mathrm{A_2S}}(t)\gamma_2\beta\sum_{j}P_j^{\mathrm{A_1I}}(t)b_{ji},
\end{align}
\end{subequations}
where $P_j^{\mathrm{A_1}}(t)=P_j^{\mathrm{A_1S}}(t)+P_j^{\mathrm{A_1I}}(t)$ and $P_i^{\mathrm{US}}(t)+P_i^{\mathrm{A_1S}}(t)+P_i^{\mathrm{A_2S}}(t)+P_i^{\mathrm{A_1I}}(t)=1$.
The fraction of infected individuals, positive information individuals, and negative information individuals is defined as $\tilde{\rho}^{\mathrm{I}}$, $\tilde{\rho}^{\mathrm{A_1}}$, and $\tilde{\rho}^{\mathrm{A_2}}$, respectively. Once the system has reached its steady state, these fractions can be calculated as follows
\begin{eqnarray}\label{eqq2}\nonumber
\rho^{\mathrm{I}}&=&\tilde{\rho}^{\mathrm{I}}(\infty)=\frac{1}{N}\sum_i P_i^{\mathrm{A_1I}}(\infty), \\ \nonumber
\rho^{\mathrm{A_1}}&=&\tilde{\rho}^{\mathrm{A_1}}(\infty)=\frac{1}{N}\sum_i \left[P_i^{\mathrm{A_1S}}(\infty)+P_i^{\mathrm{A_1I}}(\infty)\right], \\
\rho^{\mathrm{A_2}}&=&\tilde{\rho}^{\mathrm{A_2}}(\infty)=\frac{1}{N}\sum_i P_i^{\mathrm{A_2S}}(\infty).
\end{eqnarray}
Here, $N$ is network size.

Near the epidemic thresholds $\beta_c$, we set $P_i^{\mathrm{A_1I}}=\epsilon_i\ll1$ and $P_i^{\mathrm{A_2S}}=P_i^{\mathrm{A_2}}$. Consequently, $P_i^{\mathrm{A_1S}}\approx P_i^{\mathrm{A_1}}$ and $P_i^{\mathrm{US}}\approx1-P_i^{\mathrm{A_1}}-P_i^{\mathrm{A_2}}$. Inserting these and the stationarity in Eq.~\eqref{eqq1}, we obtain
\begin{subequations}\label{eqq3}
\begin{align}\label{eqq3a}
P_i^{\mathrm{A_1}}=&\frac{\lambda_1}{\delta_1}\sum_{j}\left(1-P_i^{\mathrm{A_1}}-P_i^{\mathrm{A_2}}\right)a_{ji}P_j^{\mathrm{A_1}},\\\label{eqq3b}
P_i^{\mathrm{A_2}}=&\frac{\lambda_2}{\delta_2}\sum_{j}\left(1-P_i^{\mathrm{A_1}}-P_i^{\mathrm{A_2}}\right)a_{ji}P_j^{\mathrm{A_2}},\\\label{eqq3c}
\frac{\mu}{\beta}\epsilon_i=&\sum_{j}\left[1-(1-\gamma_1)P_i^{\mathrm{A_1}}-(1-\gamma_2)P_i^{\mathrm{A_2}}\right]b_{ji}\epsilon_j.
\end{align}
\end{subequations}
Defining matrix $H$ with elements ${h_{ij} =\left[1-(1-\gamma_1)P_i^{\mathrm{A_1}}\right.}\allowbreak{\left.-(1-\gamma_2)P_i^{\mathrm{A_2}}\right]b_{ji}}$, the nontrivial solutions of Eq.~\eqref{eqq3c} are eigenvectors of $H$, whose largest real eigenvalue is equal to $\frac{\mu}{\beta}$. Then, one can obtain the epidemic threshold
\begin{equation}\label{eqq4}
\beta_c=\frac{\mu}{\Lambda_{\mathrm{max}}(H)}.
\end{equation}
Note that the matrix $H$ depends on the solutions of $P_i^{\mathrm{A_1}}$ and $P_i^{\mathrm{A_2}}$, which can be obtained by iteration of  Eqs.~\eqref{eqq3a}-\eqref{eqq3b}.

\subsection{Mean-field theory for fully connected multiplex networks}
We proceed to examine a specific network, a fully connected network, within which the individuals can be conceptualized as identical. 
In order to ensure that the rates of infection and diffusion of information are independent of the network size, we define the following parameters $\beta'=\beta (N-1)$, $\lambda_1'=\lambda_1 (N-1)$, and $\lambda_2'=\lambda_2 (N-1)$.
In this instance, according to Eq.~\eqref{eqq1}, we can obtain
\begin{subequations}\label{eqq5}
\begin{align}\label{eqq5a}
\frac{d\tilde{\rho}^{\mathrm{A_1}}}{dt}=&\tilde{\rho}^{\mathrm{US}}\lambda_1'\tilde{\rho}^{\mathrm{A_1}}-\tilde{\rho}^{\mathrm{A_1S}}\delta_1+(\tilde{\rho}^{\mathrm{US}}+\tilde{\rho}^{\mathrm{A_2}}\gamma_2)\beta'\tilde{\rho}^{\mathrm{I}},\\ \label{eqq5b}
\frac{d\tilde{\rho}^{\mathrm{A_2}}}{dt}=&\tilde{\rho}^{\mathrm{US}}\lambda_2'\tilde{\rho}^{\mathrm{A_2}}-\tilde{\rho}^{\mathrm{A_2}}\delta_2-\tilde{\rho}^{\mathrm{A_2}}\gamma_2\beta' \tilde{\rho}^{\mathrm{I}},\\\label{eqq5c}
\frac{d\tilde{\rho}^{\mathrm{I}}}{dt}=&\left(\tilde{\rho}^{\mathrm{US}}+\tilde{\rho}^{\mathrm{A_1S}}\gamma_1+\tilde{\rho}^{\mathrm{A_2}}\gamma_2\right)\beta'\tilde{\rho}^{\mathrm{I}}-\mu \tilde{\rho}^{\mathrm{I}},
\end{align}
\end{subequations}
where $\tilde{\rho}^{\mathrm{US}}=1-\tilde{\rho}^{\mathrm{A_1}}-\tilde{\rho}^{\mathrm{A_2}}$ and $\tilde{\rho}^{\mathrm{A_1S}}=\tilde{\rho}^{\mathrm{A_1}}-\tilde{\rho}^{\mathrm{I}}$.

The adjacency matrix of a fully connected network is an $N$-order matrix with diagonal elements equal to zero and other elements equal to one.
Therefore, the maximum eigenvalue of the $H$ matrix in this case is $(N-1)[1-(1-\gamma_1)\rho^{\mathrm{A_1}}-(1-\gamma_2)\rho^{\mathrm{A_2}}]$. 
Then, Eq.~\eqref{eqq4} can be written as
\begin{equation}\label{eqq6}
\beta'_c=\frac{\mu}{1-(1-\gamma_1)\rho^{\mathrm{A_1}}-(1-\gamma_2)\rho^{\mathrm{A_2}}}.
\end{equation}
In the absence of infected individuals, two pairs of steady-state solutions can be obtained from Eq.~\eqref{eqq5a} and Eq.~\eqref{eqq5b}, as follows
\begin{equation}\label{eqq7}
(\rho^{\mathrm{A_1}},\rho^{\mathrm{A_2}})=\left\{
   \begin{array}{cl}
   (1-\frac{\delta_1}{\lambda'_1},0), & \rm{for\ }\lambda'_1\delta_2>\lambda'_2\delta_1;\\
   (0,1-\frac{\delta_2}{\lambda'_2}), & \rm{for\ }\lambda'_1\delta_2<\lambda'_2\delta_1.
   \end{array}
  \right.
\end{equation}
For $\lambda'_1\delta_2=\lambda'_2\delta_1$, the sum of $\rho^{\mathrm{A_1}}$ and $\rho^{\mathrm{A_2}}$ is a fixed value $1-\frac{\delta_1}{\lambda'_1}$.

\section{Results and discussion}\label{result}
It should be noted at the outset that all discussion of $\lambda_1$, $\lambda_2$, and $\beta_c$ in Sec.~\ref{res_a} and Sec.~\ref{res_b} pertains to $\lambda'_1$, $\lambda'_2$, and $\beta'_c$ as well, given that they differ from one another by only one constant.

\subsection{Information coexistence at epidemic threshold}\label{res_a}
\begin{figure}
	\includegraphics[width=\linewidth]{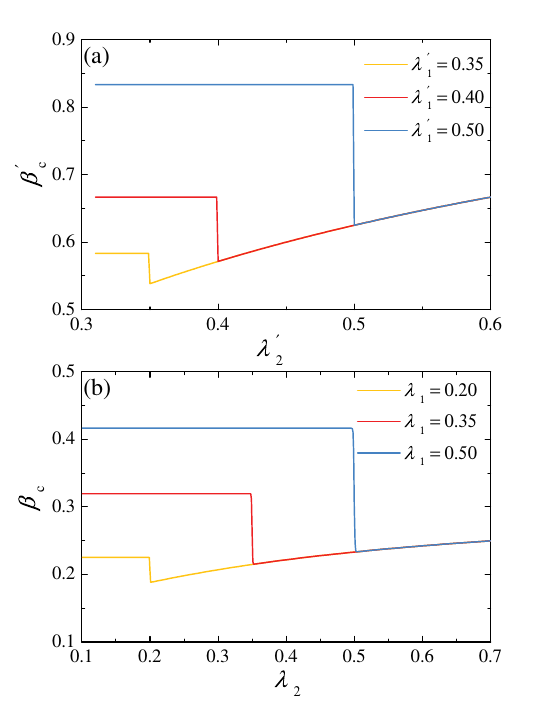}
	\caption{The epidemic threshold as a function of awareness rate of negative information for different awareness rate of positive information on mutiplex networks. (a) Both the information layer and the disease layer are fully connected networks. The results are obtained from Eq.~\eqref{eqq6}. (b) The information layer uses the BA network with an average degree of $6$, and the epidemic layer is the ER network with an average degree of $5$. The results are obtained from Eq.~\eqref{eqq4}. Parameters: $\gamma_1=0$, $\gamma_2=0.5$; (a) $\delta_1=\delta_2=0.3$, $\mu=0.5$; (b) $\delta_1=\delta_2=1$, $\mu=1$, $N=10^4$.
}\label{fig:fig2}
\end{figure}

The initial objective is to examine the impact of the awareness rates of two distinct types of information (i.e., $\lambda_1$ and $\lambda_2$) on the epidemic threshold and to discuss the coexistence of positive and negative information at the threshold.
The results were analyzed on both a fully connected network and a general network, as shown in Fig.~\ref{fig:fig2}.

Figure~\ref{fig:fig2} illustrates the epidemic threshold, which demonstrates the occurrence of a mutation at $\lambda_1=\lambda_2$. 
When $\lambda_1>\lambda_2$, $\beta_c$ increases in tandem with the rise in $\lambda_1$, exhibiting independence on $\lambda_2$. 
Conversely, when $\lambda_1<\lambda_2$, $\beta_c$ rises in proportion to the growth of $\lambda_2$, with no correlation to $\lambda_1$.
The results indicate that, in the absence of infected individuals, the population of $\mathrm{A_2}$ will extinct when $\lambda_1>\lambda_2$, while the population of $\mathrm{A_1}$ will disappear when $\lambda_1<\lambda_2$.
Meanwhile, an increase in the number of individuals who are aware of the disease (either through positive or negative preventive information) serves to impede the disease's spread and to elevate the epidemic threshold.
In the event that the infected individual is absent, the $\mathrm{UA_1A_2U}$-SIS model reverts to the $\mathrm{UA_1A_2U}$ model, thereby rendering the incompatibility between the two types of information readily apparent when $\lambda_1\neq\lambda_2$ on a fully connected network, see Eq.~\eqref{eqq7}.



\subsection{Information coexistence far from epidmeic threshold}\label{res_b}
\begin{figure}
	\includegraphics[width=\linewidth]{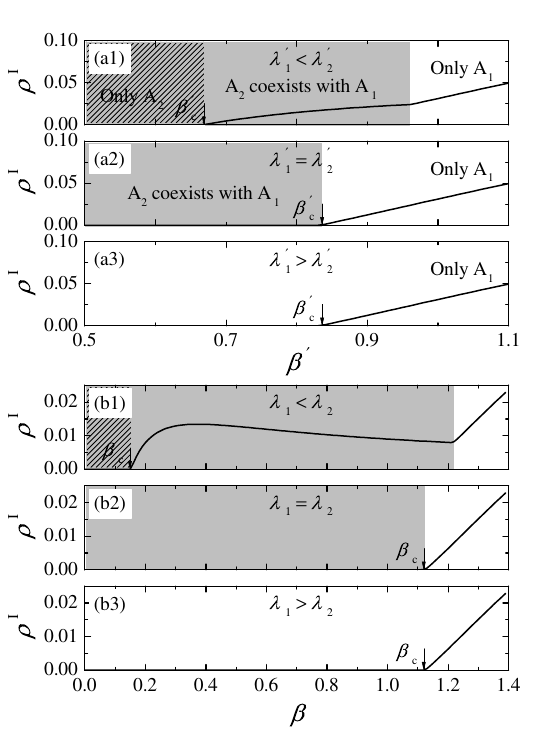}
	\caption{The epidemic prevalence $\rho^{\mathrm{I}}$ as a function of infection rate on mutiplex networks. (a) Both the information layer and the disease layer are fully connected networks. The results are obtained from Eq.~\eqref{eqq5a}-Eq.~\eqref{eqq5c}, with parameters $\delta_1=\delta_2=0.3$ and $\lambda'_1=0.5$. (b) The information layer uses the BA network with an average degree of $6$, and the epidemic layer is the ER network with an average degree of $5$. The results are obtained from Eq.~\eqref{eqq1a}-Eq.~\eqref{eqq1c}, with parameters $\delta_1=\delta_2=0.15$, $\lambda_1=0.5$, and $N=10^4$. Other parameters: $\gamma_1=0$, $\gamma_2=0.5$, $\mu=0.5$; (a1) $\lambda'_2=0.6$; (a2) $\lambda'_2=0.5$; (a3) $\lambda'_2=0.4$; (b1) $\lambda_2=0.6$; (b2) $\lambda_2=0.5$; (b3) $\lambda_2=0.4$. The diagonal shadow indicates that only $\mathrm{A_2}$ is present, without $\mathrm{A_1}$. The gray area represents the coexistence of $\mathrm{A_1}$ and $\mathrm{A_2}$ in this region. The remaining (blank) area indicates the presence of $\mathrm{A_1}$, without $\mathrm{A_2}$.
}\label{fig:fig3}
\end{figure}

Now we investigate the coexistence of positive and negative information far from the epidemic threshold and how the two distinct types of information affect the epidemic prevalence in the $\mathrm{UA_1A_2U}$-SIS model, as shown in Fig.~\ref{fig:fig3}.
As can be seen from Fig.~\ref{fig:fig3}, according to the different relationship between $\lambda_1$ and $\lambda_2$, three different phenomena will occur with the increase of $\beta$ in the population.

For $\lambda_1=\lambda_2$ and $\lambda_1>\lambda_2$, when the infected individuals are absent, a state of coexistence exists between positive and negative information in the former, whereas in the latter, only positive information is present.
Nevertheless, irrespective of whether $\lambda_1=\lambda_2$ or $\lambda_1>\lambda_2$, the $\mathrm{UA_1A_2U}$-SIS model is ultimately reduced to the classical UAU-SIS model~\cite{PhysRevLett.111.128701} in the presence of the infected individuals.

For $\lambda_1<\lambda_2$, as $\beta$ increases, there are two special points, namely a epidemic threshold point and a cusp point.
The occurrence of the epidemic threshold point indicates the emergence of infected individuals in the epidemic layer. Additionally, it marks a shift from a predominantly negative information environment to one that is more balanced, with both positive and negative information coexisting in the information layer.
The cusp point represents a shift from a context in which both positive and negative information coexist to one in which negative information is absent.
Thus, these two points are both phase transition points in the information layer.
It is noteworthy that between the epidemic threshold point and the cusp point, the density of infected individuals may demonstrate a monotonic increase or a first increase followed by a decline, as shown in Fig.~\ref{fig:fig3}(a1) and Fig.~\ref{fig:fig3}(b1).

\begin{figure}
	\includegraphics[width=\linewidth]{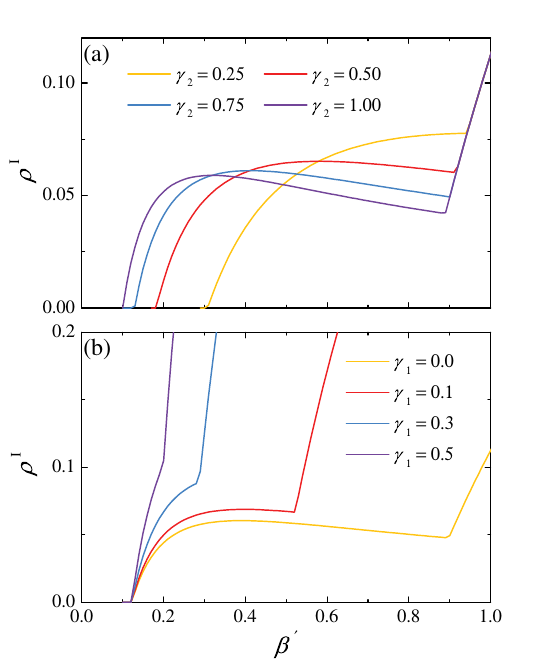}
	\caption{The epidemic prevalence $\rho^{\mathrm{I}}$ as a function of infection rate on mutiplex fully connected networks. The lines are obtained from Eq.~\eqref{eqq5a}-Eq.~\eqref{eqq5c}. Parameters: $\mu=0.1$, $\delta_1=\delta_2=0.3$, $\lambda'_1=2.5$, $\lambda'_2=3$; (a) $\gamma_1=0$; (b) $\gamma_2=0.8$.}\label{fig:fig4}
\end{figure}

It is clear that the infection rate of each individual is equal to $N\beta[1-(1-\gamma_1)\rho^{\mathrm{A_1}}-(1-\gamma_2)\rho^{\mathrm{A_2}}]$ on fully connected networks, where $\beta=\beta'/(N-1)$.
As $\beta$ increases, $\rho^{\mathrm{A_1}}$ will increase but $\rho^{\mathrm{A_2}}$ will decrease, which is contingent on the assumption that infected individuals will only appear in the $\mathrm{A_1}$I state.
Therefore, the competition between the increases of $\beta$ and $-\rho^{\mathrm{A_2}}$ and the decrease of $-\rho^{\mathrm{A_1}}$ gives rise to two different phenomena in Fig.~\ref{fig:fig3}(a1) and Fig.~\ref{fig:fig3}(b1).
Here, it can be predicted that as $\gamma_2$ increases or $\gamma_1$ decreases, the phenomenon of first rising and subsequent falling may become more pronounced.
Concurrently, the change in $\gamma_2$ does not affect the curve of the missing region of $\mathrm{A_2}$, while the change in $\gamma_1$ does.
Figure~\ref{fig:fig4}(a) shows that the phenomenon of first rising and subsequent falling becomes more significant as $\gamma_2$ increases, and all curves coincide after crossing the cusp.
Figure~\ref{fig:fig4}(b) shows that the phenomenon of first rising and subsequent falling becomes more significant as $\gamma_1$ decreases.

\subsection{Analysis of coexistence conditions and further discussion of phenomena}
To facilitate a more comprehensive analysis, we present the analytical solution for fully connected networks when $\gamma_1=0$ and $\gamma_2=1$. 
Once the system has reached its steady state and sets $\gamma_1=0$ and $\gamma_2=1$, Eq.~\eqref{eqq5a}-Eq.~\eqref{eqq5c} can be rewritten as 
\begin{subequations}\label{eqq8}
\begin{align}\nonumber
0=&\left(1-\rho^{\mathrm{A_1}}-\rho^{\mathrm{A_2}}\right)\lambda_1'\rho^{\mathrm{A_1}}-\left(\rho^{\mathrm{A_1}}-\rho^{\mathrm{I}}\right)\delta_1\\\label{eqq8a}
&+\left(1-\rho^{\mathrm{A_1}}\right)\beta'\rho^{\mathrm{I}},\\ \label{eqq8b}
0=&\left[\left(1-\rho^{\mathrm{A_1}}-\rho^{\mathrm{A_2}}\right)\lambda_2'-\delta_2-\beta' \rho^{\mathrm{I}}\right]\rho^{\mathrm{A_2}},\\\label{eqq8c}
0=&\left[\left(1-\rho^{\mathrm{A_1}}\right)\beta'-\mu\right] \rho^{\mathrm{I}}.
\end{align}
\end{subequations}
Equation~\eqref{eqq8c} allows us to readily ascertain $\rho^{\mathrm{A_1}}=1-\mu/\beta'$. We can then substitute it into both Eq.~\eqref{eqq8a} and Eq.~\eqref{eqq8b} in order to solve $\rho^{\mathrm{I}}$,
\begin{equation}\label{eqq9}
\rho^{\mathrm{I}}=\left\{
   \begin{array}{cl}
   \frac{\left(\beta'-\mu\right)\left(\lambda_2'\delta_1-\lambda_1'\delta_2\right)}{\beta'\left[\lambda_2'(\delta_1+\mu)+\lambda_1'\left(\beta'-\mu\right)\right]}, & \rm{for\ }\rho^{\mathrm{A_2}}>0;\\
	\frac{\left(\beta'-\mu\right)\left(\beta'\delta_1-\mu\lambda_1'\right)}{\beta'^2\left(\delta_1+\mu\right)}, & \rm{for\ }\rho^{\mathrm{A_2}}=0.
   \end{array}
  \right.
\end{equation}

The cusp $(\beta'_{\mathrm{cusp}}, \rho^{\mathrm{I}}_{\mathrm{cusp}})$ represents the intersection point of the two lines from the Eq.~\eqref{eqq9}.
Upon taking the limit of $\beta'\to (\beta'_{\mathrm{cusp}})^{-}$, we can insert the values of $\rho^{\mathrm{A_1}}=1-\mu/\beta'$ and $\rho^{\mathrm{A_2}}\to0$ into the Eq.~\eqref{eqq8b} to obtain
\begin{equation}\label{eqq10}
\rho^{\mathrm{I}}_{\mathrm{cusp}}=\frac{\lambda_2'}{\beta'_{\mathrm{cusp}}}\left(\frac{\mu}{\beta'_{\mathrm{cusp}}}-\frac{\delta_2}{\lambda_2'}\right).
\end{equation}
By means of simplification, $\beta'_{\mathrm{cusp}}$ can be solved by the following quadratic equation with one unknown,
\begin{align}\label{eqq11}\nonumber
&\delta_1\left(\beta'_{\mathrm{cusp}}\right)^2+\left[\delta_1\delta_2+\mu\left(\delta_2-\delta_1-\lambda_1'\right)\right]\beta'_{\mathrm{cusp}}\\
&+\mu\left(\mu\lambda_1'-\mu\lambda_2'-\delta_1\lambda_2'\right)=0.
\end{align}
In regions where $\mathrm{A_1}$ and $\mathrm{A_2}$ coexist (i.e., $\rho^{\mathrm{A_2}}>0$), the position of the maximum $(\beta'_{\mathrm{max}}, \rho^{\mathrm{I}}_{\mathrm{max}})$ can be calculated by employing the Eq.~\eqref{eqq9}.
In particular, the value of $\beta'_{\mathrm{max}}$ can be determined by $d\rho^{\mathrm{I}}/d\beta'=0$, and we have
\begin{equation}\label{eqq12}
\beta'_{\mathrm{max}}=\mu+\sqrt{\frac{\lambda_2'}{\lambda_1'}\mu\left(\mu+\delta_1\right)}.
\end{equation}
Then, Eq.~\eqref{eqq12} and Eq.~\eqref{eqq9} can be combined to ascertain the value of $\rho^{\mathrm{I}}_{\mathrm{max}}$.

It can be determined that the condition $\beta'_{\mathrm{max}}=\beta'_{\mathrm{cusp}}$ represents a critical condition. When $\beta'_{\mathrm{max}}>\beta'_{\mathrm{cusp}}$, the infection curve $\rho^{\mathrm{I}}(\beta')$ displays a monotonically increasing trend. Conversely, when $\beta'_{\mathrm{max}}$ is less than $\beta'_{\mathrm{cusp}}$, the curve first increases, then decreases, and subsequently increases again.
The slope $K$ between the maximum and the cusp, where $K=(\rho^{\mathrm{I}}_{\mathrm{cusp}}-\rho^{\mathrm{I}}_{\mathrm{max}})/(\beta'_{\mathrm{cusp}}-\beta'_{\mathrm{max}})$, is employed to assess the magnitude of the decline in the infection curve.
On the premise that the slope is negative, the smaller it is, the more significant the phenomenon of first increasing and then decreasing in the region where $\mathrm{A_1}$ and $\mathrm{A_2}$ coexist.

\begin{figure}
	\includegraphics[width=\linewidth]{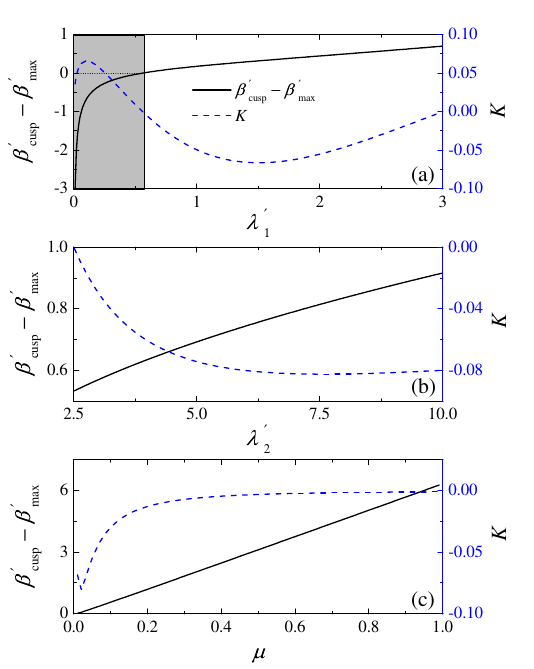}
	\caption{The criteria for evaluating the monotony of the infection curve and the rationale for determining the extent of non-monotony.
The difference $\beta'_{\mathrm{cusp}}-\beta'_{\mathrm{max}}$ and slope $K$ as a function of $\lambda'_1$ in panel (a), $\lambda'_2$ in panel (b), and $\mu$ in panel (c).
The lines and dash lines is obtained from Eq.~\eqref{eqq9}-Eq.~\eqref{eqq12}. Parameters: $\gamma_1=0$, $\gamma_2=1$, $\delta_1=\delta_2=0.3$; (a) $\mu=0.1$, $\lambda'_2=3$; (b) $\mu=0.1$, $\lambda'_1=2.5$; (c)  $\lambda'_1=2.5$, $\lambda'_2=3$.
Under the grey area parameters ($\beta'_{\mathrm{cusp}}-\beta'_{\mathrm{max}}<0$), the infected curve $\rho^{\mathrm{I}}(\beta')$ shows a monotonic increase, whereas the other regions ($\beta'_{\mathrm{cusp}}-\beta'_{\mathrm{max}}>0$) exhibit non-monotonic curves.
}\label{fig:fig5}
\end{figure}

\begin{figure}
	\includegraphics[width=\linewidth]{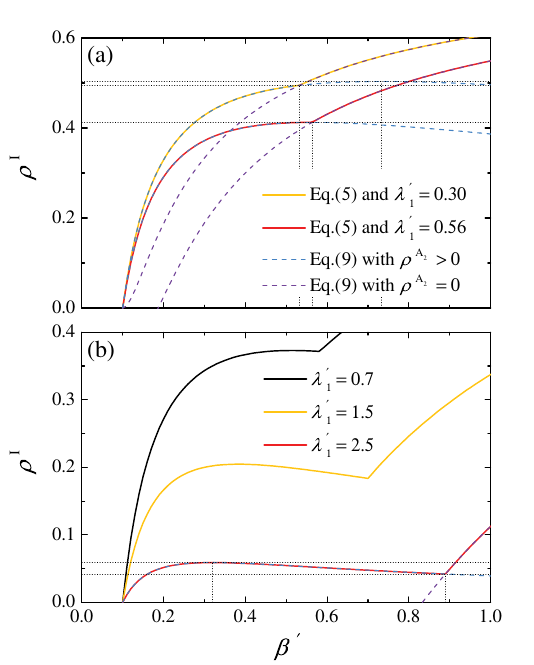}
	\caption{The epidemic prevalence $\rho^{\mathrm{I}}$ as a function of infection rate. The lines are obtained from Eq.~\eqref{eqq5}, while the dash lines are obtained from Eq.~\eqref{eqq9}. The dot lines are used for visual guidance. Parameters: $\gamma_1=0$, $\gamma_2=1$, $\delta_1=\delta_2=0.3$, $\mu=0.1$, $\lambda'_2=3$.}\label{fig:fig6}
\end{figure}

Combined with Fig.~\ref{fig:fig5} and Fig.~\ref{fig:fig6}, one learns that $\beta'_{\mathrm{cusp}}=\beta'_{\mathrm{max}}$ is the criterion for determining the monotonicity of the infection curve. 
As shown in Fig.~\ref{fig:fig6}(a), the curves display a monotonically increasing trend, with their parameters from the gray area in Fig.~\ref{fig:fig5}(a).
In contrast, the curves first increase, then decrease, and subsequently increase again in Fig.~\ref{fig:fig6}(b).

Figure~\ref{fig:fig5} shows the impact of varying $\lambda'_1$, $\lambda'_2$, and $\mu$ on the slope $K$.
Figure~\ref{fig:fig5}(a) highlights that the slope is most pronounced in the vicinity of $\lambda'_1=1.5$.
To substantiate this observation, we present Fig.~\ref{fig:fig6}(b), which shows a more precipitous decline in the infection curve when $\lambda'_1=1.5$ compared to $\lambda'_1=0.7$ and $\lambda'_1=2.5$.
As shown in Fig.~\ref{fig:fig5}(b) and Fig.~\ref{fig:fig5}(c), the slope $K$ tends towards saturation when $\lambda'_2$ and $\mu$ are relatively high.

\section{Comparison of the theory and the simulation}\label{simulation}

\begin{figure}
	\includegraphics[width=\linewidth]{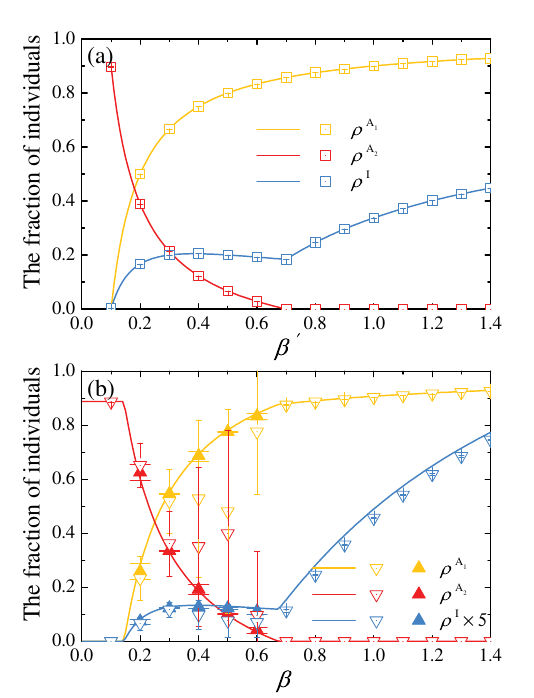}
	\caption{A comparative analysis of the simulation results with the theoretical results. The lines in panels (a) and (b) are obtained from Eq.~\eqref{eqq5} and Eq.~\eqref{eqq1}, respectively. The hollow scatters in panels (a) and (b) represent the simulation results on fully connected networks and annealed networks, respectively. The hollow scatters were obtained by averaging from 100 independent runs, with the standard deviation from the ensemble mean indicated by error bars. For $0.2\leq\beta\leq0.6$ in panel (b), the solid scatters are the surviving simulation results, obtained by removing the data with $\rho^{\mathrm{I}}<0.001$ from the original simulation results. The cap width of the error bar associated with the solid scatters is set to be wider than that of the hollow scatters. Parameters: $\gamma_1=0$, $\delta_1=\delta_2=0.3$; (a) $\gamma_2=1$, $\mu=0.1$, $\lambda'_1=1.5$, $\lambda'_2=3$; (b) $\gamma_2=0.5$, $\mu=0.5$, $\lambda_1=0.5$, $\lambda_2=0.6$.}	\label{fig:fig7}
\end{figure}

In Fig.~\ref{fig:fig7}, we present a comparison of the theoretical results with those obtained from Monte Carlo simulations.
The Gillespie algorithm~\cite{Gillespie} is employed for Monte Carlo simulations in this paper.
The duration of each simulation for fully connected networks and annealed networks is $3\times10^3$ and $2\times10^3$ units of time, respectively.
The long run equilibrium results represent the average over the last $1000$ units of time in $100$ independent simulations.
Figure~\ref{fig:fig7} demonstrates that the theoretical results are in close alignment with the simulation results.
The rationale for employing annealed networks instead of static networks, as shown in Fig.~\ref{fig:fig7}(b), is that our individual-based mean-field theory fails to account for the dynamical correlations inherent to static networks.

For the region where $\mathrm{A_1}$ and $\mathrm{A_2}$ coexist, when the number of infected individuals ($N\rho^{\mathrm{I}}$) is relatively small, the system may enter one of two absorbing states, $\rho^{\mathrm{I}}=\rho^{\mathrm{A_2}}=0$ or $\rho^{\mathrm{I}}=\rho^{\mathrm{A_1}}=0$, which is the reason for the observed fluctuation in Fig.~\ref{fig:fig7}(b).
Upon the removal of data entering the absorbing states, the simulation results and the theoretical results can be well matched with little fluctuations.

\section{conclusion}\label{conclusion}
In summary, a multiplex network is employed to investigate the impact of the co-evolution of positive and negative prevention information on epidemic thresholds and prevalence.
Here, the inhibitory effect of positive information on disease is more pronounced than that of negative information. Therefore, the presence of infected individuals facilitates the transmission of positive information.

It has been demonstrated that as the awareness rate of negative information increases, the epidemic threshold will undergo a phase transition. This phase transition occurs precisely when the awareness rate of the two kinds of information is equal.
The physical image of this phenomenon is that information with a low awareness rate will inevitably become extinct in a co-evolutionary process without external interference.
Note that our model does not facilitate the translation of the two types of information into one another.

In examining the epidemic prevalence, we focus on the scenario wherein the awareness rate of negative information is higher than that of positive information, because the remaining cases can be considered analogous to the results of the classical UAU-SIS model.
In addition to the critical epidemic threshold point, a distinctive cusp point is observed in the curve representing the density of infected individuals.
The cusp point represents a shift from a context in which both positive and negative information coexist to one in which negative information is absent.
Between the phase transition and cusp points, the density of infected individuals may exhibit a monotonic increase or a first increase followed by a decrease.
The analytical results indicate that this trend of a first increase followed by a decrease has always existed.
The physical image of this trend is that the increase of infection rate in addition to making the disease more easily spread also makes the positive information more dominant at the information layer.
However, in a realistic situation where the density is guaranteed to be non-negative, the decrease trend may not be shown up on the curve, depending on the positions of the cusp and the maximum.
Specifically, if the cusp is reached first (where the density of individuals with negative information is zero), the density of infected individuals will appear to increase monotonically.
Subsequently, the slope between between the maximum and the cusp is defined, and the downward trend is examined as each parameter is altered.

The theoretical results are then compared with those obtained from Monte Carlo simulations, including the cases of identical individuals and an arbitrary degree distribution. The results demonstrate a high degree of consistency, thus validating the theoretical approach.

In our model, the existence of negative information is contingent upon two conditions: first, that its awareness rate exceeds that of positive information; and second, that the density of infected individuals is not high. 
This is consistent with popular perception.
To better explore this issue, more assumptions may be unlocked in future studies, such as allowing the two types of information to be switched to each other.

\section*{ACKNOWLEDGEMENTS}
This work was supported by the Shaanxi Fundamental Science Research Project for Mathematics and Physics (Grant No. 22JSQ003). Si-Ping Zhang was supported by the Natural Science Basic Research Program of Shaanxi (Grant No. 2021JQ-007) and the China Postdoctoral Science Foundation (Grant No. 2020M673363).


\bibliography{document}

\begin{thebibliography}{56}%
\makeatletter
\providecommand \@ifxundefined [1]{%
 \@ifx{#1\undefined}
}%
\providecommand \@ifnum [1]{%
 \ifnum #1\expandafter \@firstoftwo
 \else \expandafter \@secondoftwo
 \fi
}%
\providecommand \@ifx [1]{%
 \ifx #1\expandafter \@firstoftwo
 \else \expandafter \@secondoftwo
 \fi
}%
\providecommand \natexlab [1]{#1}%
\providecommand \enquote  [1]{``#1''}%
\providecommand \bibnamefont  [1]{#1}%
\providecommand \bibfnamefont [1]{#1}%
\providecommand \citenamefont [1]{#1}%
\providecommand \href@noop [0]{\@secondoftwo}%
\providecommand \href [0]{\begingroup \@sanitize@url \@href}%
\providecommand \@href[1]{\@@startlink{#1}\@@href}%
\providecommand \@@href[1]{\endgroup#1\@@endlink}%
\providecommand \@sanitize@url [0]{\catcode `\\12\catcode `\$12\catcode
  `\&12\catcode `\#12\catcode `\^12\catcode `\_12\catcode `\%12\relax}%
\providecommand \@@startlink[1]{}%
\providecommand \@@endlink[0]{}%
\providecommand \url  [0]{\begingroup\@sanitize@url \@url }%
\providecommand \@url [1]{\endgroup\@href {#1}{\urlprefix }}%
\providecommand \urlprefix  [0]{URL }%
\providecommand \Eprint [0]{\href }%
\providecommand \doibase [0]{https://doi.org/}%
\providecommand \selectlanguage [0]{\@gobble}%
\providecommand \bibinfo  [0]{\@secondoftwo}%
\providecommand \bibfield  [0]{\@secondoftwo}%
\providecommand \translation [1]{[#1]}%
\providecommand \BibitemOpen [0]{}%
\providecommand \bibitemStop [0]{}%
\providecommand \bibitemNoStop [0]{.\EOS\space}%
\providecommand \EOS [0]{\spacefactor3000\relax}%
\providecommand \BibitemShut  [1]{\csname bibitem#1\endcsname}%
\let\auto@bib@innerbib\@empty
\bibitem [{\citenamefont {Scheufele}\ and\ \citenamefont
  {Krause}(2019)}]{doi:10.1073/pnas.1805871115}%
  \BibitemOpen
  \bibfield  {author} {\bibinfo {author} {\bibfnamefont {D.~A.}\ \bibnamefont
  {Scheufele}}\ and\ \bibinfo {author} {\bibfnamefont {N.~M.}\ \bibnamefont
  {Krause}},\ }\bibfield  {title} {\bibinfo {title} {Science audiences,
  misinformation, and fake news},\ }\href
  {https://doi.org/10.1073/pnas.1805871115} {\bibfield  {journal} {\bibinfo
  {journal} {Proc. Natl. Acad. Sci. U. S. A.}\ }\textbf {\bibinfo {volume}
  {116}},\ \bibinfo {pages} {7662} (\bibinfo {year} {2019})}\BibitemShut
  {NoStop}%
\bibitem [{\citenamefont {Trpevski}\ \emph {et~al.}(2010)\citenamefont
  {Trpevski}, \citenamefont {Tang},\ and\ \citenamefont
  {Kocarev}}]{PhysRevE.81.056102}%
  \BibitemOpen
  \bibfield  {author} {\bibinfo {author} {\bibfnamefont {D.}~\bibnamefont
  {Trpevski}}, \bibinfo {author} {\bibfnamefont {W.~K.~S.}\ \bibnamefont
  {Tang}},\ and\ \bibinfo {author} {\bibfnamefont {L.}~\bibnamefont
  {Kocarev}},\ }\bibfield  {title} {\bibinfo {title} {Model for rumor spreading
  over networks},\ }\href {https://doi.org/10.1103/PhysRevE.81.056102}
  {\bibfield  {journal} {\bibinfo  {journal} {Phys. Rev. E}\ }\textbf {\bibinfo
  {volume} {81}},\ \bibinfo {pages} {056102} (\bibinfo {year}
  {2010})}\BibitemShut {NoStop}%
\bibitem [{\citenamefont {Wang}\ \emph {et~al.}(2012)\citenamefont {Wang},
  \citenamefont {Xiao},\ and\ \citenamefont {Liu}}]{Wang_2012}%
  \BibitemOpen
  \bibfield  {author} {\bibinfo {author} {\bibfnamefont {Y.}~\bibnamefont
  {Wang}}, \bibinfo {author} {\bibfnamefont {G.}~\bibnamefont {Xiao}},\ and\
  \bibinfo {author} {\bibfnamefont {J.}~\bibnamefont {Liu}},\ }\bibfield
  {title} {\bibinfo {title} {Dynamics of competing ideas in complex social
  systems},\ }\href {https://doi.org/10.1088/1367-2630/14/1/013015} {\bibfield
  {journal} {\bibinfo  {journal} {New J. Phys.}\ }\textbf {\bibinfo {volume}
  {14}},\ \bibinfo {pages} {013015} (\bibinfo {year} {2012})}\BibitemShut
  {NoStop}%
\bibitem [{\citenamefont {Burghardt}\ \emph {et~al.}(2016)\citenamefont
  {Burghardt}, \citenamefont {Rand},\ and\ \citenamefont
  {Girvan}}]{PhysRevE.93.032305}%
  \BibitemOpen
  \bibfield  {author} {\bibinfo {author} {\bibfnamefont {K.}~\bibnamefont
  {Burghardt}}, \bibinfo {author} {\bibfnamefont {W.}~\bibnamefont {Rand}},\
  and\ \bibinfo {author} {\bibfnamefont {M.}~\bibnamefont {Girvan}},\
  }\bibfield  {title} {\bibinfo {title} {Competing opinions and stubborness:
  Connecting models to data},\ }\href
  {https://doi.org/10.1103/PhysRevE.93.032305} {\bibfield  {journal} {\bibinfo
  {journal} {Phys. Rev. E}\ }\textbf {\bibinfo {volume} {93}},\ \bibinfo
  {pages} {032305} (\bibinfo {year} {2016})}\BibitemShut {NoStop}%
\bibitem [{\citenamefont {Wang}\ and\ \citenamefont {Zhao}(2017)}]{WANG201763}%
  \BibitemOpen
  \bibfield  {author} {\bibinfo {author} {\bibfnamefont {X.}~\bibnamefont
  {Wang}}\ and\ \bibinfo {author} {\bibfnamefont {T.}~\bibnamefont {Zhao}},\
  }\bibfield  {title} {\bibinfo {title} {Model for multi-messages spreading
  over complex networks considering the relationship between messages},\ }\href
  {https://doi.org/https://doi.org/10.1016/j.cnsns.2016.12.019} {\bibfield
  {journal} {\bibinfo  {journal} {Commun. Nonlinear Sci. Numer. Simul.}\
  }\textbf {\bibinfo {volume} {48}},\ \bibinfo {pages} {63} (\bibinfo {year}
  {2017})}\BibitemShut {NoStop}%
\bibitem [{\citenamefont {Vosoughi}\ \emph {et~al.}(2018)\citenamefont
  {Vosoughi}, \citenamefont {Roy},\ and\ \citenamefont
  {Aral}}]{doi:10.1126/science.aap9559}%
  \BibitemOpen
  \bibfield  {author} {\bibinfo {author} {\bibfnamefont {S.}~\bibnamefont
  {Vosoughi}}, \bibinfo {author} {\bibfnamefont {D.}~\bibnamefont {Roy}},\ and\
  \bibinfo {author} {\bibfnamefont {S.}~\bibnamefont {Aral}},\ }\bibfield
  {title} {\bibinfo {title} {The spread of true and false news online},\ }\href
  {https://doi.org/10.1126/science.aap9559} {\bibfield  {journal} {\bibinfo
  {journal} {Science}\ }\textbf {\bibinfo {volume} {359}},\ \bibinfo {pages}
  {1146} (\bibinfo {year} {2018})}\BibitemShut {NoStop}%
\bibitem [{\citenamefont {Lazer}\ \emph {et~al.}(2018)\citenamefont {Lazer},
  \citenamefont {Baum}, \citenamefont {Benkler}, \citenamefont {Berinsky},
  \citenamefont {Greenhill}, \citenamefont {Menczer}, \citenamefont {Metzger},
  \citenamefont {Nyhan}, \citenamefont {Pennycook}, \citenamefont {Rothschild},
  \citenamefont {Schudson}, \citenamefont {Sloman}, \citenamefont {Sunstein},
  \citenamefont {Thorson}, \citenamefont {Watts},\ and\ \citenamefont
  {Zittrain}}]{doi:10.1126/science.aao2998}%
  \BibitemOpen
  \bibfield  {author} {\bibinfo {author} {\bibfnamefont {D.~M.~J.}\
  \bibnamefont {Lazer}}, \bibinfo {author} {\bibfnamefont {M.~A.}\ \bibnamefont
  {Baum}}, \bibinfo {author} {\bibfnamefont {Y.}~\bibnamefont {Benkler}},
  \bibinfo {author} {\bibfnamefont {A.~J.}\ \bibnamefont {Berinsky}}, \bibinfo
  {author} {\bibfnamefont {K.~M.}\ \bibnamefont {Greenhill}}, \bibinfo {author}
  {\bibfnamefont {F.}~\bibnamefont {Menczer}}, \bibinfo {author} {\bibfnamefont
  {M.~J.}\ \bibnamefont {Metzger}}, \bibinfo {author} {\bibfnamefont
  {B.}~\bibnamefont {Nyhan}}, \bibinfo {author} {\bibfnamefont
  {G.}~\bibnamefont {Pennycook}}, \bibinfo {author} {\bibfnamefont
  {D.}~\bibnamefont {Rothschild}}, \bibinfo {author} {\bibfnamefont
  {M.}~\bibnamefont {Schudson}}, \bibinfo {author} {\bibfnamefont {S.~A.}\
  \bibnamefont {Sloman}}, \bibinfo {author} {\bibfnamefont {C.~R.}\
  \bibnamefont {Sunstein}}, \bibinfo {author} {\bibfnamefont {E.~A.}\
  \bibnamefont {Thorson}}, \bibinfo {author} {\bibfnamefont {D.~J.}\
  \bibnamefont {Watts}},\ and\ \bibinfo {author} {\bibfnamefont {J.~L.}\
  \bibnamefont {Zittrain}},\ }\bibfield  {title} {\bibinfo {title} {The science
  of fake news},\ }\href {https://doi.org/10.1126/science.aao2998} {\bibfield
  {journal} {\bibinfo  {journal} {Science}\ }\textbf {\bibinfo {volume}
  {359}},\ \bibinfo {pages} {1094} (\bibinfo {year} {2018})}\BibitemShut
  {NoStop}%
\bibitem [{\citenamefont {Baumann}\ \emph {et~al.}(2020)\citenamefont
  {Baumann}, \citenamefont {Lorenz-Spreen}, \citenamefont {Sokolov},\ and\
  \citenamefont {Starnini}}]{PhysRevLett.124.048301}%
  \BibitemOpen
  \bibfield  {author} {\bibinfo {author} {\bibfnamefont {F.}~\bibnamefont
  {Baumann}}, \bibinfo {author} {\bibfnamefont {P.}~\bibnamefont
  {Lorenz-Spreen}}, \bibinfo {author} {\bibfnamefont {I.~M.}\ \bibnamefont
  {Sokolov}},\ and\ \bibinfo {author} {\bibfnamefont {M.}~\bibnamefont
  {Starnini}},\ }\bibfield  {title} {\bibinfo {title} {Modeling echo chambers
  and polarization dynamics in social networks},\ }\href
  {https://doi.org/10.1103/PhysRevLett.124.048301} {\bibfield  {journal}
  {\bibinfo  {journal} {Phys. Rev. Lett.}\ }\textbf {\bibinfo {volume} {124}},\
  \bibinfo {pages} {048301} (\bibinfo {year} {2020})}\BibitemShut {NoStop}%
\bibitem [{\citenamefont {Yao}\ \emph {et~al.}(2020)\citenamefont {Yao},
  \citenamefont {Li}, \citenamefont {Xiong}, \citenamefont {Wu}, \citenamefont
  {Lin},\ and\ \citenamefont {Ju}}]{YAO2020122764}%
  \BibitemOpen
  \bibfield  {author} {\bibinfo {author} {\bibfnamefont {Y.}~\bibnamefont
  {Yao}}, \bibinfo {author} {\bibfnamefont {Y.}~\bibnamefont {Li}}, \bibinfo
  {author} {\bibfnamefont {X.}~\bibnamefont {Xiong}}, \bibinfo {author}
  {\bibfnamefont {Y.}~\bibnamefont {Wu}}, \bibinfo {author} {\bibfnamefont
  {H.}~\bibnamefont {Lin}},\ and\ \bibinfo {author} {\bibfnamefont
  {S.}~\bibnamefont {Ju}},\ }\bibfield  {title} {\bibinfo {title} {An
  interactive propagation model of multiple information in complex networks},\
  }\href {https://doi.org/https://doi.org/10.1016/j.physa.2019.122764}
  {\bibfield  {journal} {\bibinfo  {journal} {Physica A}\ }\textbf {\bibinfo
  {volume} {537}},\ \bibinfo {pages} {122764} (\bibinfo {year}
  {2020})}\BibitemShut {NoStop}%
\bibitem [{\citenamefont {Zhang}\ \emph {et~al.}(2022)\citenamefont {Zhang},
  \citenamefont {Chen}, \citenamefont {Peng}, \citenamefont {Kou},\ and\
  \citenamefont {Wang}}]{ZHANG2022366}%
  \BibitemOpen
  \bibfield  {author} {\bibinfo {author} {\bibfnamefont {H.}~\bibnamefont
  {Zhang}}, \bibinfo {author} {\bibfnamefont {X.}~\bibnamefont {Chen}},
  \bibinfo {author} {\bibfnamefont {Y.}~\bibnamefont {Peng}}, \bibinfo {author}
  {\bibfnamefont {G.}~\bibnamefont {Kou}},\ and\ \bibinfo {author}
  {\bibfnamefont {R.}~\bibnamefont {Wang}},\ }\bibfield  {title} {\bibinfo
  {title} {The interaction of multiple information on multiplex social
  networks},\ }\href
  {https://doi.org/https://doi.org/10.1016/j.ins.2022.05.036} {\bibfield
  {journal} {\bibinfo  {journal} {Inf. Sci.}\ }\textbf {\bibinfo {volume}
  {605}},\ \bibinfo {pages} {366} (\bibinfo {year} {2022})}\BibitemShut
  {NoStop}%
\bibitem [{\citenamefont {Gajewski}\ \emph {et~al.}(2022)\citenamefont
  {Gajewski}, \citenamefont {Sienkiewicz},\ and\ \citenamefont
  {Ho\l{}yst}}]{PhysRevE.105.024125}%
  \BibitemOpen
  \bibfield  {author} {\bibinfo {author} {\bibfnamefont {L.~G.}\ \bibnamefont
  {Gajewski}}, \bibinfo {author} {\bibfnamefont {J.}~\bibnamefont
  {Sienkiewicz}},\ and\ \bibinfo {author} {\bibfnamefont {J.~A.}\ \bibnamefont
  {Ho\l{}yst}},\ }\bibfield  {title} {\bibinfo {title} {Transitions between
  polarization and radicalization in a temporal bilayer echo-chamber model},\
  }\href {https://doi.org/10.1103/PhysRevE.105.024125} {\bibfield  {journal}
  {\bibinfo  {journal} {Phys. Rev. E}\ }\textbf {\bibinfo {volume} {105}},\
  \bibinfo {pages} {024125} (\bibinfo {year} {2022})}\BibitemShut {NoStop}%
\bibitem [{\citenamefont {Chiyomaru}\ and\ \citenamefont
  {Takemoto}(2022)}]{PhysRevE.106.014301}%
  \BibitemOpen
  \bibfield  {author} {\bibinfo {author} {\bibfnamefont {K.}~\bibnamefont
  {Chiyomaru}}\ and\ \bibinfo {author} {\bibfnamefont {K.}~\bibnamefont
  {Takemoto}},\ }\bibfield  {title} {\bibinfo {title} {Adversarial attacks on
  voter model dynamics in complex networks},\ }\href
  {https://doi.org/10.1103/PhysRevE.106.014301} {\bibfield  {journal} {\bibinfo
   {journal} {Phys. Rev. E}\ }\textbf {\bibinfo {volume} {106}},\ \bibinfo
  {pages} {014301} (\bibinfo {year} {2022})}\BibitemShut {NoStop}%
\bibitem [{\citenamefont {Wang}\ \emph {et~al.}(2019)\citenamefont {Wang},
  \citenamefont {Wang}, \citenamefont {Hao}, \citenamefont {Min},\ and\
  \citenamefont {Wang}}]{8717734}%
  \BibitemOpen
  \bibfield  {author} {\bibinfo {author} {\bibfnamefont {X.}~\bibnamefont
  {Wang}}, \bibinfo {author} {\bibfnamefont {X.}~\bibnamefont {Wang}}, \bibinfo
  {author} {\bibfnamefont {F.}~\bibnamefont {Hao}}, \bibinfo {author}
  {\bibfnamefont {G.}~\bibnamefont {Min}},\ and\ \bibinfo {author}
  {\bibfnamefont {L.}~\bibnamefont {Wang}},\ }\bibfield  {title} {\bibinfo
  {title} {Efficient coupling diffusion of positive and negative information in
  online social networks},\ }\href {https://doi.org/10.1109/TNSM.2019.2917512}
  {\bibfield  {journal} {\bibinfo  {journal} {IEEE Trans. Netw. Serv. Manag.}\
  }\textbf {\bibinfo {volume} {16}},\ \bibinfo {pages} {1226} (\bibinfo {year}
  {2019})}\BibitemShut {NoStop}%
\bibitem [{\citenamefont {Wang}\ \emph
  {et~al.}(2022{\natexlab{a}})\citenamefont {Wang}, \citenamefont {Wang},
  \citenamefont {Min}, \citenamefont {Hao},\ and\ \citenamefont
  {Chen}}]{9036077}%
  \BibitemOpen
  \bibfield  {author} {\bibinfo {author} {\bibfnamefont {X.}~\bibnamefont
  {Wang}}, \bibinfo {author} {\bibfnamefont {X.}~\bibnamefont {Wang}}, \bibinfo
  {author} {\bibfnamefont {G.}~\bibnamefont {Min}}, \bibinfo {author}
  {\bibfnamefont {F.}~\bibnamefont {Hao}},\ and\ \bibinfo {author}
  {\bibfnamefont {C.~L.~P.}\ \bibnamefont {Chen}},\ }\bibfield  {title}
  {\bibinfo {title} {An efficient feedback control mechanism for
  positive/negative information spread in online social networks},\ }\href
  {https://doi.org/10.1109/TCYB.2020.2977322} {\bibfield  {journal} {\bibinfo
  {journal} {IEEE T. Cybern.}\ }\textbf {\bibinfo {volume} {52}},\ \bibinfo
  {pages} {87} (\bibinfo {year} {2022}{\natexlab{a}})}\BibitemShut {NoStop}%
\bibitem [{\citenamefont {Shi}\ \emph {et~al.}(2024)\citenamefont {Shi},
  \citenamefont {Chen}, \citenamefont {Zhong},\ and\ \citenamefont
  {Zhang}}]{10336945}%
  \BibitemOpen
  \bibfield  {author} {\bibinfo {author} {\bibfnamefont {X.-L.}\ \bibnamefont
  {Shi}}, \bibinfo {author} {\bibfnamefont {W.-N.}\ \bibnamefont {Chen}},
  \bibinfo {author} {\bibfnamefont {J.-H.}\ \bibnamefont {Zhong}},\ and\
  \bibinfo {author} {\bibfnamefont {J.}~\bibnamefont {Zhang}},\ }\bibfield
  {title} {\bibinfo {title} {A max–min ant system with repetitive influence
  reduction strategy for interactive dissemination of positive and negative
  information},\ }\href {https://doi.org/10.1109/TCSS.2023.3328994} {\bibfield
  {journal} {\bibinfo  {journal} {IEEE Trans. Comput. Soc. Syst.}\ }\textbf
  {\bibinfo {volume} {11}},\ \bibinfo {pages} {3255} (\bibinfo {year}
  {2024})}\BibitemShut {NoStop}%
\bibitem [{\citenamefont {Chu}\ \emph {et~al.}(2024)\citenamefont {Chu},
  \citenamefont {Song}, \citenamefont {Zhao}, \citenamefont {Chen},\ and\
  \citenamefont {Chiang}}]{Chu2024}%
  \BibitemOpen
  \bibfield  {author} {\bibinfo {author} {\bibfnamefont {M.}~\bibnamefont
  {Chu}}, \bibinfo {author} {\bibfnamefont {W.}~\bibnamefont {Song}}, \bibinfo
  {author} {\bibfnamefont {Z.}~\bibnamefont {Zhao}}, \bibinfo {author}
  {\bibfnamefont {T.}~\bibnamefont {Chen}},\ and\ \bibinfo {author}
  {\bibfnamefont {Y.-c.}\ \bibnamefont {Chiang}},\ }\bibfield  {title}
  {\bibinfo {title} {Emotional contagion on social media and the simulation of
  intervention strategies after a disaster event: a modeling study},\ }\href
  {https://doi.org/10.1057/s41599-024-03397-4} {\bibfield  {journal} {\bibinfo
  {journal} {Hum. Soc. Sci. Commun.}\ }\textbf {\bibinfo {volume} {11}},\
  \bibinfo {pages} {968} (\bibinfo {year} {2024})}\BibitemShut {NoStop}%
\bibitem [{\citenamefont {Kobayashi}(2022)}]{PhysRevE.106.034303}%
  \BibitemOpen
  \bibfield  {author} {\bibinfo {author} {\bibfnamefont {T.}~\bibnamefont
  {Kobayashi}},\ }\bibfield  {title} {\bibinfo {title} {Diffusion dynamics of
  competing information on networks},\ }\href
  {https://doi.org/10.1103/PhysRevE.106.034303} {\bibfield  {journal} {\bibinfo
   {journal} {Phys. Rev. E}\ }\textbf {\bibinfo {volume} {106}},\ \bibinfo
  {pages} {034303} (\bibinfo {year} {2022})}\BibitemShut {NoStop}%
\bibitem [{\citenamefont {Wan}\ \emph {et~al.}(2023)\citenamefont {Wan},
  \citenamefont {Wang}, \citenamefont {Min}, \citenamefont {Wang},
  \citenamefont {Lin}, \citenamefont {Yu},\ and\ \citenamefont {Wu}}]{9915429}%
  \BibitemOpen
  \bibfield  {author} {\bibinfo {author} {\bibfnamefont {P.}~\bibnamefont
  {Wan}}, \bibinfo {author} {\bibfnamefont {X.}~\bibnamefont {Wang}}, \bibinfo
  {author} {\bibfnamefont {G.}~\bibnamefont {Min}}, \bibinfo {author}
  {\bibfnamefont {L.}~\bibnamefont {Wang}}, \bibinfo {author} {\bibfnamefont
  {Y.}~\bibnamefont {Lin}}, \bibinfo {author} {\bibfnamefont {W.}~\bibnamefont
  {Yu}},\ and\ \bibinfo {author} {\bibfnamefont {X.}~\bibnamefont {Wu}},\
  }\bibfield  {title} {\bibinfo {title} {Optimal control for positive and
  negative information diffusion based on game theory in online social
  networks},\ }\href {https://doi.org/10.1109/TNSE.2022.3211988} {\bibfield
  {journal} {\bibinfo  {journal} {IEEE Trans. Netw. Sci. Eng.}\ }\textbf
  {\bibinfo {volume} {10}},\ \bibinfo {pages} {426} (\bibinfo {year}
  {2023})}\BibitemShut {NoStop}%
\bibitem [{\citenamefont {Goldenberg}\ \emph {et~al.}(2007)\citenamefont
  {Goldenberg}, \citenamefont {Libai}, \citenamefont {Moldovan},\ and\
  \citenamefont {Muller}}]{GOLDENBERG2007186}%
  \BibitemOpen
  \bibfield  {author} {\bibinfo {author} {\bibfnamefont {J.}~\bibnamefont
  {Goldenberg}}, \bibinfo {author} {\bibfnamefont {B.}~\bibnamefont {Libai}},
  \bibinfo {author} {\bibfnamefont {S.}~\bibnamefont {Moldovan}},\ and\
  \bibinfo {author} {\bibfnamefont {E.}~\bibnamefont {Muller}},\ }\bibfield
  {title} {\bibinfo {title} {{The NPV of bad news}},\ }\href
  {https://doi.org/https://doi.org/10.1016/j.ijresmar.2007.02.003} {\bibfield
  {journal} {\bibinfo  {journal} {Int. J. Res. Mark.}\ }\textbf {\bibinfo
  {volume} {24}},\ \bibinfo {pages} {186} (\bibinfo {year} {2007})}\BibitemShut
  {NoStop}%
\bibitem [{\citenamefont {Loomba}\ \emph {et~al.}(2021)\citenamefont {Loomba},
  \citenamefont {De~Figueiredo}, \citenamefont {Piatek}, \citenamefont
  {De~Graaf},\ and\ \citenamefont {Larson}}]{loomba2021measuring}%
  \BibitemOpen
  \bibfield  {author} {\bibinfo {author} {\bibfnamefont {S.}~\bibnamefont
  {Loomba}}, \bibinfo {author} {\bibfnamefont {A.}~\bibnamefont
  {De~Figueiredo}}, \bibinfo {author} {\bibfnamefont {S.~J.}\ \bibnamefont
  {Piatek}}, \bibinfo {author} {\bibfnamefont {K.}~\bibnamefont {De~Graaf}},\
  and\ \bibinfo {author} {\bibfnamefont {H.~J.}\ \bibnamefont {Larson}},\
  }\bibfield  {title} {\bibinfo {title} {{Measuring the impact of COVID-19
  vaccine misinformation on vaccination intent in the UK and USA}},\ }\href
  {https://doi.org/10.1038/s41562-021-01056-1} {\bibfield  {journal} {\bibinfo
  {journal} {Nat. Hum. Behav.}\ }\textbf {\bibinfo {volume} {5}},\ \bibinfo
  {pages} {337} (\bibinfo {year} {2021})}\BibitemShut {NoStop}%
\bibitem [{\citenamefont {Allen}\ \emph {et~al.}(2024)\citenamefont {Allen},
  \citenamefont {Watts},\ and\ \citenamefont
  {Rand}}]{doi:10.1126/science.adk3451}%
  \BibitemOpen
  \bibfield  {author} {\bibinfo {author} {\bibfnamefont {J.}~\bibnamefont
  {Allen}}, \bibinfo {author} {\bibfnamefont {D.~J.}\ \bibnamefont {Watts}},\
  and\ \bibinfo {author} {\bibfnamefont {D.~G.}\ \bibnamefont {Rand}},\
  }\bibfield  {title} {\bibinfo {title} {{Quantifying the impact of
  misinformation and vaccine-skeptical content on Facebook}},\ }\href
  {https://doi.org/10.1126/science.adk3451} {\bibfield  {journal} {\bibinfo
  {journal} {Science}\ }\textbf {\bibinfo {volume} {384}},\ \bibinfo {pages}
  {eadk3451} (\bibinfo {year} {2024})}\BibitemShut {NoStop}%
\bibitem [{\citenamefont {Moehring}\ \emph {et~al.}(2023)\citenamefont
  {Moehring}, \citenamefont {Collis}, \citenamefont {Garimella}, \citenamefont
  {Rahimian}, \citenamefont {Aral},\ and\ \citenamefont
  {Eckles}}]{Moehring2023}%
  \BibitemOpen
  \bibfield  {author} {\bibinfo {author} {\bibfnamefont {A.}~\bibnamefont
  {Moehring}}, \bibinfo {author} {\bibfnamefont {A.}~\bibnamefont {Collis}},
  \bibinfo {author} {\bibfnamefont {K.}~\bibnamefont {Garimella}}, \bibinfo
  {author} {\bibfnamefont {M.~A.}\ \bibnamefont {Rahimian}}, \bibinfo {author}
  {\bibfnamefont {S.}~\bibnamefont {Aral}},\ and\ \bibinfo {author}
  {\bibfnamefont {D.}~\bibnamefont {Eckles}},\ }\bibfield  {title} {\bibinfo
  {title} {{Providing normative information increases intentions to accept a
  COVID-19 vaccine}},\ }\href {https://doi.org/10.1038/s41467-022-35052-4}
  {\bibfield  {journal} {\bibinfo  {journal} {Nat. Commun.}\ }\textbf {\bibinfo
  {volume} {14}},\ \bibinfo {pages} {126} (\bibinfo {year} {2023})}\BibitemShut
  {NoStop}%
\bibitem [{\citenamefont {Ferguson}(2007)}]{Ferguson2007}%
  \BibitemOpen
  \bibfield  {author} {\bibinfo {author} {\bibfnamefont {N.}~\bibnamefont
  {Ferguson}},\ }\bibfield  {title} {\bibinfo {title} {Capturing human
  behaviour},\ }\href {https://doi.org/10.1038/446733a} {\bibfield  {journal}
  {\bibinfo  {journal} {Nature}\ }\textbf {\bibinfo {volume} {446}},\ \bibinfo
  {pages} {733} (\bibinfo {year} {2007})}\BibitemShut {NoStop}%
\bibitem [{\citenamefont {Wang}\ \emph {et~al.}(2015)\citenamefont {Wang},
  \citenamefont {Andrews}, \citenamefont {Wu}, \citenamefont {Wang},\ and\
  \citenamefont {Bauch}}]{WANG20151}%
  \BibitemOpen
  \bibfield  {author} {\bibinfo {author} {\bibfnamefont {Z.}~\bibnamefont
  {Wang}}, \bibinfo {author} {\bibfnamefont {M.~A.}\ \bibnamefont {Andrews}},
  \bibinfo {author} {\bibfnamefont {Z.-X.}\ \bibnamefont {Wu}}, \bibinfo
  {author} {\bibfnamefont {L.}~\bibnamefont {Wang}},\ and\ \bibinfo {author}
  {\bibfnamefont {C.~T.}\ \bibnamefont {Bauch}},\ }\bibfield  {title} {\bibinfo
  {title} {Coupled disease-behavior dynamics on complex networks: A review},\
  }\href {https://doi.org/10.1016/j.plrev.2015.07.006} {\bibfield  {journal}
  {\bibinfo  {journal} {Phys. Life Rev.}\ }\textbf {\bibinfo {volume} {15}},\
  \bibinfo {pages} {1} (\bibinfo {year} {2015})}\BibitemShut {NoStop}%
\bibitem [{\citenamefont {Granell}\ \emph {et~al.}(2013)\citenamefont
  {Granell}, \citenamefont {G\'omez},\ and\ \citenamefont
  {Arenas}}]{PhysRevLett.111.128701}%
  \BibitemOpen
  \bibfield  {author} {\bibinfo {author} {\bibfnamefont {C.}~\bibnamefont
  {Granell}}, \bibinfo {author} {\bibfnamefont {S.}~\bibnamefont {G\'omez}},\
  and\ \bibinfo {author} {\bibfnamefont {A.}~\bibnamefont {Arenas}},\
  }\bibfield  {title} {\bibinfo {title} {Dynamical interplay between awareness
  and epidemic spreading in multiplex networks},\ }\href
  {https://doi.org/10.1103/PhysRevLett.111.128701} {\bibfield  {journal}
  {\bibinfo  {journal} {Phys. Rev. Lett.}\ }\textbf {\bibinfo {volume} {111}},\
  \bibinfo {pages} {128701} (\bibinfo {year} {2013})}\BibitemShut {NoStop}%
\bibitem [{\citenamefont {Granell}\ \emph {et~al.}(2014)\citenamefont
  {Granell}, \citenamefont {G\'omez},\ and\ \citenamefont
  {Arenas}}]{PhysRevE.90.012808}%
  \BibitemOpen
  \bibfield  {author} {\bibinfo {author} {\bibfnamefont {C.}~\bibnamefont
  {Granell}}, \bibinfo {author} {\bibfnamefont {S.}~\bibnamefont {G\'omez}},\
  and\ \bibinfo {author} {\bibfnamefont {A.}~\bibnamefont {Arenas}},\
  }\bibfield  {title} {\bibinfo {title} {Competing spreading processes on
  multiplex networks: Awareness and epidemics},\ }\href
  {https://doi.org/10.1103/PhysRevE.90.012808} {\bibfield  {journal} {\bibinfo
  {journal} {Phys. Rev. E}\ }\textbf {\bibinfo {volume} {90}},\ \bibinfo
  {pages} {012808} (\bibinfo {year} {2014})}\BibitemShut {NoStop}%
\bibitem [{\citenamefont {Chang}\ \emph {et~al.}(2021)\citenamefont {Chang},
  \citenamefont {Cai}, \citenamefont {Zhang},\ and\ \citenamefont
  {Wang}}]{PhysRevE.104.044303}%
  \BibitemOpen
  \bibfield  {author} {\bibinfo {author} {\bibfnamefont {X.}~\bibnamefont
  {Chang}}, \bibinfo {author} {\bibfnamefont {C.-R.}\ \bibnamefont {Cai}},
  \bibinfo {author} {\bibfnamefont {J.-Q.}\ \bibnamefont {Zhang}},\ and\
  \bibinfo {author} {\bibfnamefont {C.-Y.}\ \bibnamefont {Wang}},\ }\bibfield
  {title} {\bibinfo {title} {Analytical solution of epidemic threshold for
  coupled information-epidemic dynamics on multiplex networks with alterable
  heterogeneity},\ }\href {https://doi.org/10.1103/PhysRevE.104.044303}
  {\bibfield  {journal} {\bibinfo  {journal} {Phys. Rev. E}\ }\textbf {\bibinfo
  {volume} {104}},\ \bibinfo {pages} {044303} (\bibinfo {year}
  {2021})}\BibitemShut {NoStop}%
\bibitem [{\citenamefont {Chang}\ \emph {et~al.}(2023)\citenamefont {Chang},
  \citenamefont {Cai}, \citenamefont {Wang}, \citenamefont {Liu}, \citenamefont
  {Zhang}, \citenamefont {Jin},\ and\ \citenamefont
  {Yang}}]{PhysRevResearch.5.013196}%
  \BibitemOpen
  \bibfield  {author} {\bibinfo {author} {\bibfnamefont {X.}~\bibnamefont
  {Chang}}, \bibinfo {author} {\bibfnamefont {C.-R.}\ \bibnamefont {Cai}},
  \bibinfo {author} {\bibfnamefont {C.-Y.}\ \bibnamefont {Wang}}, \bibinfo
  {author} {\bibfnamefont {X.-S.}\ \bibnamefont {Liu}}, \bibinfo {author}
  {\bibfnamefont {J.-Q.}\ \bibnamefont {Zhang}}, \bibinfo {author}
  {\bibfnamefont {K.}~\bibnamefont {Jin}},\ and\ \bibinfo {author}
  {\bibfnamefont {W.-L.}\ \bibnamefont {Yang}},\ }\bibfield  {title} {\bibinfo
  {title} {Combined effect of simplicial complexes and interlayer interaction:
  An example of information-epidemic dynamics on multiplex networks},\ }\href
  {https://doi.org/10.1103/PhysRevResearch.5.013196} {\bibfield  {journal}
  {\bibinfo  {journal} {Phys. Rev. Res.}\ }\textbf {\bibinfo {volume} {5}},\
  \bibinfo {pages} {013196} (\bibinfo {year} {2023})}\BibitemShut {NoStop}%
\bibitem [{\citenamefont {Zhou}\ \emph {et~al.}(2019)\citenamefont {Zhou},
  \citenamefont {Zhou}, \citenamefont {Chen},\ and\ \citenamefont
  {Stanley}}]{Zhou_2019}%
  \BibitemOpen
  \bibfield  {author} {\bibinfo {author} {\bibfnamefont {Y.}~\bibnamefont
  {Zhou}}, \bibinfo {author} {\bibfnamefont {J.}~\bibnamefont {Zhou}}, \bibinfo
  {author} {\bibfnamefont {G.}~\bibnamefont {Chen}},\ and\ \bibinfo {author}
  {\bibfnamefont {H.~E.}\ \bibnamefont {Stanley}},\ }\bibfield  {title}
  {\bibinfo {title} {Effective degree theory for awareness and epidemic
  spreading on multiplex networks},\ }\href
  {https://doi.org/10.1088/1367-2630/ab0458} {\bibfield  {journal} {\bibinfo
  {journal} {New J. Phys.}\ }\textbf {\bibinfo {volume} {21}},\ \bibinfo
  {pages} {035002} (\bibinfo {year} {2019})}\BibitemShut {NoStop}%
\bibitem [{\citenamefont {da~Silva}\ \emph {et~al.}(2019)\citenamefont
  {da~Silva}, \citenamefont {Vel\'asquez-Rojas}, \citenamefont {Connaughton},
  \citenamefont {Vazquez}, \citenamefont {Moreno},\ and\ \citenamefont
  {Rodrigues}}]{PhysRevE.100.032313}%
  \BibitemOpen
  \bibfield  {author} {\bibinfo {author} {\bibfnamefont {P.~C.~V.}\
  \bibnamefont {da~Silva}}, \bibinfo {author} {\bibfnamefont {F.}~\bibnamefont
  {Vel\'asquez-Rojas}}, \bibinfo {author} {\bibfnamefont {C.}~\bibnamefont
  {Connaughton}}, \bibinfo {author} {\bibfnamefont {F.}~\bibnamefont
  {Vazquez}}, \bibinfo {author} {\bibfnamefont {Y.}~\bibnamefont {Moreno}},\
  and\ \bibinfo {author} {\bibfnamefont {F.~A.}\ \bibnamefont {Rodrigues}},\
  }\bibfield  {title} {\bibinfo {title} {Epidemic spreading with awareness and
  different timescales in multiplex networks},\ }\href
  {https://doi.org/10.1103/PhysRevE.100.032313} {\bibfield  {journal} {\bibinfo
   {journal} {Phys. Rev. E}\ }\textbf {\bibinfo {volume} {100}},\ \bibinfo
  {pages} {032313} (\bibinfo {year} {2019})}\BibitemShut {NoStop}%
\bibitem [{\citenamefont {Vel\'asquez-Rojas}\ \emph {et~al.}(2020)\citenamefont
  {Vel\'asquez-Rojas}, \citenamefont {Ventura}, \citenamefont {Connaughton},
  \citenamefont {Moreno}, \citenamefont {Rodrigues},\ and\ \citenamefont
  {Vazquez}}]{PhysRevE.102.022312}%
  \BibitemOpen
  \bibfield  {author} {\bibinfo {author} {\bibfnamefont {F.}~\bibnamefont
  {Vel\'asquez-Rojas}}, \bibinfo {author} {\bibfnamefont {P.~C.}\ \bibnamefont
  {Ventura}}, \bibinfo {author} {\bibfnamefont {C.}~\bibnamefont
  {Connaughton}}, \bibinfo {author} {\bibfnamefont {Y.}~\bibnamefont {Moreno}},
  \bibinfo {author} {\bibfnamefont {F.~A.}\ \bibnamefont {Rodrigues}},\ and\
  \bibinfo {author} {\bibfnamefont {F.}~\bibnamefont {Vazquez}},\ }\bibfield
  {title} {\bibinfo {title} {Disease and information spreading at different
  speeds in multiplex networks},\ }\href
  {https://doi.org/10.1103/PhysRevE.102.022312} {\bibfield  {journal} {\bibinfo
   {journal} {Phys. Rev. E}\ }\textbf {\bibinfo {volume} {102}},\ \bibinfo
  {pages} {022312} (\bibinfo {year} {2020})}\BibitemShut {NoStop}%
\bibitem [{\citenamefont {Kabir}\ \emph {et~al.}(2020)\citenamefont {Kabir},
  \citenamefont {Kuga},\ and\ \citenamefont {Tanimoto}}]{kabir2020impact}%
  \BibitemOpen
  \bibfield  {author} {\bibinfo {author} {\bibfnamefont {K.~A.}\ \bibnamefont
  {Kabir}}, \bibinfo {author} {\bibfnamefont {K.}~\bibnamefont {Kuga}},\ and\
  \bibinfo {author} {\bibfnamefont {J.}~\bibnamefont {Tanimoto}},\ }\bibfield
  {title} {\bibinfo {title} {{The impact of information spreading on epidemic
  vaccination game dynamics in a heterogeneous complex network- A theoretical
  approach}},\ }\href
  {https://doi.org/https://doi.org/10.1016/j.chaos.2019.109548} {\bibfield
  {journal} {\bibinfo  {journal} {Chaos Solitons Fractals}\ }\textbf {\bibinfo
  {volume} {132}},\ \bibinfo {pages} {109548} (\bibinfo {year}
  {2020})}\BibitemShut {NoStop}%
\bibitem [{\citenamefont {Guo}\ \emph {et~al.}(2022)\citenamefont {Guo},
  \citenamefont {Tu}, \citenamefont {Shen},\ and\ \citenamefont
  {Chai}}]{PhysRevE.106.034307}%
  \BibitemOpen
  \bibfield  {author} {\bibinfo {author} {\bibfnamefont {Y.}~\bibnamefont
  {Guo}}, \bibinfo {author} {\bibfnamefont {L.}~\bibnamefont {Tu}}, \bibinfo
  {author} {\bibfnamefont {H.}~\bibnamefont {Shen}},\ and\ \bibinfo {author}
  {\bibfnamefont {L.}~\bibnamefont {Chai}},\ }\bibfield  {title} {\bibinfo
  {title} {Transmission dynamics of disease spreading in multilayer networks
  with mass media},\ }\href {https://doi.org/10.1103/PhysRevE.106.034307}
  {\bibfield  {journal} {\bibinfo  {journal} {Phys. Rev. E}\ }\textbf {\bibinfo
  {volume} {106}},\ \bibinfo {pages} {034307} (\bibinfo {year}
  {2022})}\BibitemShut {NoStop}%
\bibitem [{\citenamefont {Ye}\ \emph {et~al.}(2020)\citenamefont {Ye},
  \citenamefont {Zhang}, \citenamefont {Ruan}, \citenamefont {Cao},
  \citenamefont {Xuan},\ and\ \citenamefont {Zeng}}]{PhysRevE.102.042314}%
  \BibitemOpen
  \bibfield  {author} {\bibinfo {author} {\bibfnamefont {Y.}~\bibnamefont
  {Ye}}, \bibinfo {author} {\bibfnamefont {Q.}~\bibnamefont {Zhang}}, \bibinfo
  {author} {\bibfnamefont {Z.}~\bibnamefont {Ruan}}, \bibinfo {author}
  {\bibfnamefont {Z.}~\bibnamefont {Cao}}, \bibinfo {author} {\bibfnamefont
  {Q.}~\bibnamefont {Xuan}},\ and\ \bibinfo {author} {\bibfnamefont {D.~D.}\
  \bibnamefont {Zeng}},\ }\bibfield  {title} {\bibinfo {title} {Effect of
  heterogeneous risk perception on information diffusion, behavior change, and
  disease transmission},\ }\href {https://doi.org/10.1103/PhysRevE.102.042314}
  {\bibfield  {journal} {\bibinfo  {journal} {Phys. Rev. E}\ }\textbf {\bibinfo
  {volume} {102}},\ \bibinfo {pages} {042314} (\bibinfo {year}
  {2020})}\BibitemShut {NoStop}%
\bibitem [{\citenamefont {Wang}\ \emph
  {et~al.}(2021{\natexlab{a}})\citenamefont {Wang}, \citenamefont {Zhu},
  \citenamefont {Tao}, \citenamefont {Xiao}, \citenamefont {Wang},\ and\
  \citenamefont {Lai}}]{PhysRevResearch.3.013157}%
  \BibitemOpen
  \bibfield  {author} {\bibinfo {author} {\bibfnamefont {X.}~\bibnamefont
  {Wang}}, \bibinfo {author} {\bibfnamefont {X.}~\bibnamefont {Zhu}}, \bibinfo
  {author} {\bibfnamefont {X.}~\bibnamefont {Tao}}, \bibinfo {author}
  {\bibfnamefont {J.}~\bibnamefont {Xiao}}, \bibinfo {author} {\bibfnamefont
  {W.}~\bibnamefont {Wang}},\ and\ \bibinfo {author} {\bibfnamefont {Y.-C.}\
  \bibnamefont {Lai}},\ }\bibfield  {title} {\bibinfo {title} {Anomalous role
  of information diffusion in epidemic spreading},\ }\href
  {https://doi.org/10.1103/PhysRevResearch.3.013157} {\bibfield  {journal}
  {\bibinfo  {journal} {Phys. Rev. Res.}\ }\textbf {\bibinfo {volume} {3}},\
  \bibinfo {pages} {013157} (\bibinfo {year} {2021}{\natexlab{a}})}\BibitemShut
  {NoStop}%
\bibitem [{\citenamefont {Guo}\ \emph {et~al.}(2021{\natexlab{a}})\citenamefont
  {Guo}, \citenamefont {Wang}, \citenamefont {Sun},\ and\ \citenamefont
  {Xia}}]{GUO2021127282}%
  \BibitemOpen
  \bibfield  {author} {\bibinfo {author} {\bibfnamefont {H.}~\bibnamefont
  {Guo}}, \bibinfo {author} {\bibfnamefont {Z.}~\bibnamefont {Wang}}, \bibinfo
  {author} {\bibfnamefont {S.}~\bibnamefont {Sun}},\ and\ \bibinfo {author}
  {\bibfnamefont {C.}~\bibnamefont {Xia}},\ }\bibfield  {title} {\bibinfo
  {title} {Interplay between epidemic spread and information diffusion on
  two-layered networks with partial mapping},\ }\href
  {https://doi.org/https://doi.org/10.1016/j.physleta.2021.127282} {\bibfield
  {journal} {\bibinfo  {journal} {Phys. Lett. A}\ }\textbf {\bibinfo {volume}
  {398}},\ \bibinfo {pages} {127282} (\bibinfo {year}
  {2021}{\natexlab{a}})}\BibitemShut {NoStop}%
\bibitem [{\citenamefont {Guo}\ \emph {et~al.}(2021{\natexlab{b}})\citenamefont
  {Guo}, \citenamefont {Yin}, \citenamefont {Xia},\ and\ \citenamefont
  {Dehmer}}]{guo2021impact}%
  \BibitemOpen
  \bibfield  {author} {\bibinfo {author} {\bibfnamefont {H.}~\bibnamefont
  {Guo}}, \bibinfo {author} {\bibfnamefont {Q.}~\bibnamefont {Yin}}, \bibinfo
  {author} {\bibfnamefont {C.}~\bibnamefont {Xia}},\ and\ \bibinfo {author}
  {\bibfnamefont {M.}~\bibnamefont {Dehmer}},\ }\bibfield  {title} {\bibinfo
  {title} {Impact of information diffusion on epidemic spreading in partially
  mapping two-layered time-varying networks},\ }\href
  {https://doi.org/10.1007/s11071-021-06784-7} {\bibfield  {journal} {\bibinfo
  {journal} {Nonlinear Dyn.}\ }\textbf {\bibinfo {volume} {105}},\ \bibinfo
  {pages} {3819} (\bibinfo {year} {2021}{\natexlab{b}})}\BibitemShut {NoStop}%
\bibitem [{\citenamefont {Wang}\ \emph
  {et~al.}(2022{\natexlab{b}})\citenamefont {Wang}, \citenamefont {Zhang},
  \citenamefont {Zhu},\ and\ \citenamefont {Ma}}]{10.1063/5.0099183}%
  \BibitemOpen
  \bibfield  {author} {\bibinfo {author} {\bibfnamefont {H.}~\bibnamefont
  {Wang}}, \bibinfo {author} {\bibfnamefont {H.-F.}\ \bibnamefont {Zhang}},
  \bibinfo {author} {\bibfnamefont {P.-C.}\ \bibnamefont {Zhu}},\ and\ \bibinfo
  {author} {\bibfnamefont {C.}~\bibnamefont {Ma}},\ }\bibfield  {title}
  {\bibinfo {title} {Interplay of simplicial awareness contagion and epidemic
  spreading on time-varying multiplex networks},\ }\href
  {https://doi.org/10.1063/5.0099183} {\bibfield  {journal} {\bibinfo
  {journal} {Chaos}\ }\textbf {\bibinfo {volume} {32}},\ \bibinfo {pages}
  {083110} (\bibinfo {year} {2022}{\natexlab{b}})}\BibitemShut {NoStop}%
\bibitem [{\citenamefont {Liu}\ \emph {et~al.}(2023)\citenamefont {Liu},
  \citenamefont {Feng}, \citenamefont {Xia}, \citenamefont {Zhao},\ and\
  \citenamefont {Perc}}]{LIU2023113657}%
  \BibitemOpen
  \bibfield  {author} {\bibinfo {author} {\bibfnamefont {L.}~\bibnamefont
  {Liu}}, \bibinfo {author} {\bibfnamefont {M.}~\bibnamefont {Feng}}, \bibinfo
  {author} {\bibfnamefont {C.}~\bibnamefont {Xia}}, \bibinfo {author}
  {\bibfnamefont {D.}~\bibnamefont {Zhao}},\ and\ \bibinfo {author}
  {\bibfnamefont {M.}~\bibnamefont {Perc}},\ }\bibfield  {title} {\bibinfo
  {title} {Epidemic trajectories and awareness diffusion among unequals in
  simplicial complexes},\ }\href
  {https://doi.org/https://doi.org/10.1016/j.chaos.2023.113657} {\bibfield
  {journal} {\bibinfo  {journal} {Chaos Solitons Fractals}\ }\textbf {\bibinfo
  {volume} {173}},\ \bibinfo {pages} {113657} (\bibinfo {year}
  {2023})}\BibitemShut {NoStop}%
\bibitem [{\citenamefont {Yin}\ \emph {et~al.}(2023)\citenamefont {Yin},
  \citenamefont {Wang},\ and\ \citenamefont {Xia}}]{Yin2023}%
  \BibitemOpen
  \bibfield  {author} {\bibinfo {author} {\bibfnamefont {Q.}~\bibnamefont
  {Yin}}, \bibinfo {author} {\bibfnamefont {Z.}~\bibnamefont {Wang}},\ and\
  \bibinfo {author} {\bibfnamefont {C.}~\bibnamefont {Xia}},\ }\bibfield
  {title} {\bibinfo {title} {Information-epidemic co-evolution propagation
  under policy intervention in multiplex networks},\ }\href
  {https://doi.org/10.1007/s11071-023-08581-w} {\bibfield  {journal} {\bibinfo
  {journal} {Nonlinear Dyn.}\ }\textbf {\bibinfo {volume} {111}},\ \bibinfo
  {pages} {14583} (\bibinfo {year} {2023})}\BibitemShut {NoStop}%
\bibitem [{\citenamefont {Cai}\ \emph {et~al.}(2023)\citenamefont {Cai},
  \citenamefont {Liu}, \citenamefont {Chang},\ and\ \citenamefont
  {Liu}}]{PhysRevResearch.5.033220}%
  \BibitemOpen
  \bibfield  {author} {\bibinfo {author} {\bibfnamefont {C.-R.}\ \bibnamefont
  {Cai}}, \bibinfo {author} {\bibfnamefont {N.-N.}\ \bibnamefont {Liu}},
  \bibinfo {author} {\bibfnamefont {X.}~\bibnamefont {Chang}},\ and\ \bibinfo
  {author} {\bibfnamefont {X.-S.}\ \bibnamefont {Liu}},\ }\bibfield  {title}
  {\bibinfo {title} {Physical images of relative timescales in coevolution
  dynamics},\ }\href {https://doi.org/10.1103/PhysRevResearch.5.033220}
  {\bibfield  {journal} {\bibinfo  {journal} {Phys. Rev. Res.}\ }\textbf
  {\bibinfo {volume} {5}},\ \bibinfo {pages} {033220} (\bibinfo {year}
  {2023})}\BibitemShut {NoStop}%
\bibitem [{\citenamefont {Masoomy}\ \emph {et~al.}(2023)\citenamefont
  {Masoomy}, \citenamefont {Chou},\ and\ \citenamefont
  {Böttcher}}]{Masoomy_2023}%
  \BibitemOpen
  \bibfield  {author} {\bibinfo {author} {\bibfnamefont {H.}~\bibnamefont
  {Masoomy}}, \bibinfo {author} {\bibfnamefont {T.}~\bibnamefont {Chou}},\ and\
  \bibinfo {author} {\bibfnamefont {L.}~\bibnamefont {Böttcher}},\ }\bibfield
  {title} {\bibinfo {title} {Impact of random and targeted disruptions on
  information diffusion during outbreaks},\ }\href
  {https://doi.org/10.1063/5.0139844} {\bibfield  {journal} {\bibinfo
  {journal} {Chaos}\ }\textbf {\bibinfo {volume} {33}},\ \bibinfo {pages}
  {033145} (\bibinfo {year} {2023})}\BibitemShut {NoStop}%
\bibitem [{\citenamefont {Chen}\ \emph {et~al.}(2023)\citenamefont {Chen},
  \citenamefont {Hu},\ and\ \citenamefont {Cao}}]{PhysRevResearch.5.033065}%
  \BibitemOpen
  \bibfield  {author} {\bibinfo {author} {\bibfnamefont {J.}~\bibnamefont
  {Chen}}, \bibinfo {author} {\bibfnamefont {M.}~\bibnamefont {Hu}},\ and\
  \bibinfo {author} {\bibfnamefont {J.}~\bibnamefont {Cao}},\ }\bibfield
  {title} {\bibinfo {title} {Dynamics of
  information-awareness-epidemic-activity coevolution in multiplex networks},\
  }\href {https://doi.org/10.1103/PhysRevResearch.5.033065} {\bibfield
  {journal} {\bibinfo  {journal} {Phys. Rev. Res.}\ }\textbf {\bibinfo {volume}
  {5}},\ \bibinfo {pages} {033065} (\bibinfo {year} {2023})}\BibitemShut
  {NoStop}%
\bibitem [{\citenamefont {Chang}\ \emph {et~al.}(2024)\citenamefont {Chang},
  \citenamefont {Cai}, \citenamefont {Zhang},\ and\ \citenamefont
  {Yang}}]{CHANG2024114780}%
  \BibitemOpen
  \bibfield  {author} {\bibinfo {author} {\bibfnamefont {X.}~\bibnamefont
  {Chang}}, \bibinfo {author} {\bibfnamefont {C.-R.}\ \bibnamefont {Cai}},
  \bibinfo {author} {\bibfnamefont {J.-Q.}\ \bibnamefont {Zhang}},\ and\
  \bibinfo {author} {\bibfnamefont {W.-L.}\ \bibnamefont {Yang}},\ }\bibfield
  {title} {\bibinfo {title} {The universality of physical images at relative
  timescales on multiplex networks},\ }\href
  {https://doi.org/https://doi.org/10.1016/j.chaos.2024.114780} {\bibfield
  {journal} {\bibinfo  {journal} {Chaos Solitons Fractals}\ }\textbf {\bibinfo
  {volume} {182}},\ \bibinfo {pages} {114780} (\bibinfo {year}
  {2024})}\BibitemShut {NoStop}%
\bibitem [{\citenamefont {Gallotti}\ \emph {et~al.}(2020)\citenamefont
  {Gallotti}, \citenamefont {Valle}, \citenamefont {Castaldo}, \citenamefont
  {Sacco},\ and\ \citenamefont {De~Domenico}}]{gallotti2020assessing}%
  \BibitemOpen
  \bibfield  {author} {\bibinfo {author} {\bibfnamefont {R.}~\bibnamefont
  {Gallotti}}, \bibinfo {author} {\bibfnamefont {F.}~\bibnamefont {Valle}},
  \bibinfo {author} {\bibfnamefont {N.}~\bibnamefont {Castaldo}}, \bibinfo
  {author} {\bibfnamefont {P.}~\bibnamefont {Sacco}},\ and\ \bibinfo {author}
  {\bibfnamefont {M.}~\bibnamefont {De~Domenico}},\ }\bibfield  {title}
  {\bibinfo {title} {{Assessing the risks of ‘infodemics’ in response to
  COVID-19 epidemics}},\ }\href {https://doi.org/10.1038/s41562-020-00994-6}
  {\bibfield  {journal} {\bibinfo  {journal} {Nat. Hum. Behav.}\ }\textbf
  {\bibinfo {volume} {4}},\ \bibinfo {pages} {1285} (\bibinfo {year}
  {2020})}\BibitemShut {NoStop}%
\bibitem [{\citenamefont {Wang}\ \emph
  {et~al.}(2021{\natexlab{b}})\citenamefont {Wang}, \citenamefont {Xia},
  \citenamefont {Chen},\ and\ \citenamefont {Chen}}]{8957067}%
  \BibitemOpen
  \bibfield  {author} {\bibinfo {author} {\bibfnamefont {Z.}~\bibnamefont
  {Wang}}, \bibinfo {author} {\bibfnamefont {C.}~\bibnamefont {Xia}}, \bibinfo
  {author} {\bibfnamefont {Z.}~\bibnamefont {Chen}},\ and\ \bibinfo {author}
  {\bibfnamefont {G.}~\bibnamefont {Chen}},\ }\bibfield  {title} {\bibinfo
  {title} {Epidemic propagation with positive and negative preventive
  information in multiplex networks},\ }\href
  {https://doi.org/10.1109/TCYB.2019.2960605} {\bibfield  {journal} {\bibinfo
  {journal} {IEEE T. Cybern.}\ }\textbf {\bibinfo {volume} {51}},\ \bibinfo
  {pages} {1454} (\bibinfo {year} {2021}{\natexlab{b}})}\BibitemShut {NoStop}%
\bibitem [{\citenamefont {Huang}\ \emph {et~al.}(2021)\citenamefont {Huang},
  \citenamefont {Chen},\ and\ \citenamefont {Ma}}]{HUANG2021125536}%
  \BibitemOpen
  \bibfield  {author} {\bibinfo {author} {\bibfnamefont {H.}~\bibnamefont
  {Huang}}, \bibinfo {author} {\bibfnamefont {Y.}~\bibnamefont {Chen}},\ and\
  \bibinfo {author} {\bibfnamefont {Y.}~\bibnamefont {Ma}},\ }\bibfield
  {title} {\bibinfo {title} {Modeling the competitive diffusions of rumor and
  knowledge and the impacts on epidemic spreading},\ }\href
  {https://doi.org/https://doi.org/10.1016/j.amc.2020.125536} {\bibfield
  {journal} {\bibinfo  {journal} {Appl. Math. Comput.}\ }\textbf {\bibinfo
  {volume} {388}},\ \bibinfo {pages} {125536} (\bibinfo {year}
  {2021})}\BibitemShut {NoStop}%
\bibitem [{\citenamefont {Chen}\ \emph {et~al.}(2022)\citenamefont {Chen},
  \citenamefont {Liu}, \citenamefont {Yue}, \citenamefont {Duan},\ and\
  \citenamefont {Tang}}]{Chen2022CoevolvingSD}%
  \BibitemOpen
  \bibfield  {author} {\bibinfo {author} {\bibfnamefont {J.}~\bibnamefont
  {Chen}}, \bibinfo {author} {\bibfnamefont {Y.}~\bibnamefont {Liu}}, \bibinfo
  {author} {\bibfnamefont {J.}~\bibnamefont {Yue}}, \bibinfo {author}
  {\bibfnamefont {X.}~\bibnamefont {Duan}},\ and\ \bibinfo {author}
  {\bibfnamefont {M.}~\bibnamefont {Tang}},\ }\bibfield  {title} {\bibinfo
  {title} {Coevolving spreading dynamics of negative information and epidemic
  on multiplex networks},\ }\href
  {https://api.semanticscholar.org/CorpusID:251764681} {\bibfield  {journal}
  {\bibinfo  {journal} {Nonlinear Dyn.}\ }\textbf {\bibinfo {volume} {110}},\
  \bibinfo {pages} {3881 } (\bibinfo {year} {2022})}\BibitemShut {NoStop}%
\bibitem [{\citenamefont {Wu}\ and\ \citenamefont
  {Bao}(2022)}]{10.1063/5.0126799}%
  \BibitemOpen
  \bibfield  {author} {\bibinfo {author} {\bibfnamefont {X.}~\bibnamefont
  {Wu}}\ and\ \bibinfo {author} {\bibfnamefont {H.}~\bibnamefont {Bao}},\
  }\bibfield  {title} {\bibinfo {title} {{The impact of positive and negative
  information on SIR-like epidemics in delayed multiplex networks}},\ }\href
  {https://doi.org/10.1063/5.0126799} {\bibfield  {journal} {\bibinfo
  {journal} {Chaos}\ }\textbf {\bibinfo {volume} {32}},\ \bibinfo {pages}
  {113141} (\bibinfo {year} {2022})}\BibitemShut {NoStop}%
\bibitem [{\citenamefont {Fang}\ \emph {et~al.}(2023)\citenamefont {Fang},
  \citenamefont {Ma},\ and\ \citenamefont {Li}}]{FANG2023113376}%
  \BibitemOpen
  \bibfield  {author} {\bibinfo {author} {\bibfnamefont {F.}~\bibnamefont
  {Fang}}, \bibinfo {author} {\bibfnamefont {J.}~\bibnamefont {Ma}},\ and\
  \bibinfo {author} {\bibfnamefont {Y.}~\bibnamefont {Li}},\ }\bibfield
  {title} {\bibinfo {title} {The coevolution of the spread of a disease and
  competing opinions in multiplex networks},\ }\href
  {https://doi.org/https://doi.org/10.1016/j.chaos.2023.113376} {\bibfield
  {journal} {\bibinfo  {journal} {Chaos Solitons Fractals}\ }\textbf {\bibinfo
  {volume} {170}},\ \bibinfo {pages} {113376} (\bibinfo {year}
  {2023})}\BibitemShut {NoStop}%
\bibitem [{\citenamefont {Li}\ \emph {et~al.}(2024)\citenamefont {Li},
  \citenamefont {Xie},\ and\ \citenamefont {Han}}]{LI2024128700}%
  \BibitemOpen
  \bibfield  {author} {\bibinfo {author} {\bibfnamefont {D.}~\bibnamefont
  {Li}}, \bibinfo {author} {\bibfnamefont {W.}~\bibnamefont {Xie}},\ and\
  \bibinfo {author} {\bibfnamefont {D.}~\bibnamefont {Han}},\ }\bibfield
  {title} {\bibinfo {title} {Multi-information and epidemic coupling
  propagation considering indirect contact on two-layer networks},\ }\href
  {https://doi.org/https://doi.org/10.1016/j.amc.2024.128700} {\bibfield
  {journal} {\bibinfo  {journal} {Appl. Math. Comput.}\ }\textbf {\bibinfo
  {volume} {474}},\ \bibinfo {pages} {128700} (\bibinfo {year}
  {2024})}\BibitemShut {NoStop}%
\bibitem [{\citenamefont {Han}\ and\ \citenamefont
  {Wang}(2024)}]{HAN2024115264}%
  \BibitemOpen
  \bibfield  {author} {\bibinfo {author} {\bibfnamefont {D.}~\bibnamefont
  {Han}}\ and\ \bibinfo {author} {\bibfnamefont {X.}~\bibnamefont {Wang}},\
  }\bibfield  {title} {\bibinfo {title} {Impact of positive and negative
  information on epidemic spread in a three-layer network},\ }\href
  {https://doi.org/https://doi.org/10.1016/j.chaos.2024.115264} {\bibfield
  {journal} {\bibinfo  {journal} {Chaos Solitons Fractals}\ }\textbf {\bibinfo
  {volume} {186}},\ \bibinfo {pages} {115264} (\bibinfo {year}
  {2024})}\BibitemShut {NoStop}%
\bibitem [{\citenamefont {Pastor-Satorras}\ \emph {et~al.}(2015)\citenamefont
  {Pastor-Satorras}, \citenamefont {Castellano}, \citenamefont {Van~Mieghem},\
  and\ \citenamefont {Vespignani}}]{pastor2015epidemic}%
  \BibitemOpen
  \bibfield  {author} {\bibinfo {author} {\bibfnamefont {R.}~\bibnamefont
  {Pastor-Satorras}}, \bibinfo {author} {\bibfnamefont {C.}~\bibnamefont
  {Castellano}}, \bibinfo {author} {\bibfnamefont {P.}~\bibnamefont
  {Van~Mieghem}},\ and\ \bibinfo {author} {\bibfnamefont {A.}~\bibnamefont
  {Vespignani}},\ }\bibfield  {title} {\bibinfo {title} {Epidemic processes in
  complex networks},\ }\href {https://doi.org/10.1103/RevModPhys.87.925}
  {\bibfield  {journal} {\bibinfo  {journal} {Rev. Mod. Phys.}\ }\textbf
  {\bibinfo {volume} {87}},\ \bibinfo {pages} {925} (\bibinfo {year}
  {2015})}\BibitemShut {NoStop}%
\bibitem [{\citenamefont {Cai}\ \emph {et~al.}(2016)\citenamefont {Cai},
  \citenamefont {Wu}, \citenamefont {Chen}, \citenamefont {Holme},\ and\
  \citenamefont {Guan}}]{PhysRevLett.116.258301}%
  \BibitemOpen
  \bibfield  {author} {\bibinfo {author} {\bibfnamefont {C.-R.}\ \bibnamefont
  {Cai}}, \bibinfo {author} {\bibfnamefont {Z.-X.}\ \bibnamefont {Wu}},
  \bibinfo {author} {\bibfnamefont {M.~Z.~Q.}\ \bibnamefont {Chen}}, \bibinfo
  {author} {\bibfnamefont {P.}~\bibnamefont {Holme}},\ and\ \bibinfo {author}
  {\bibfnamefont {J.-Y.}\ \bibnamefont {Guan}},\ }\bibfield  {title} {\bibinfo
  {title} {Solving the dynamic correlation problem of the
  susceptible-infected-susceptible model on networks},\ }\href
  {https://doi.org/10.1103/PhysRevLett.116.258301} {\bibfield  {journal}
  {\bibinfo  {journal} {Phys. Rev. Lett.}\ }\textbf {\bibinfo {volume} {116}},\
  \bibinfo {pages} {258301} (\bibinfo {year} {2016})}\BibitemShut {NoStop}%
\bibitem [{\citenamefont {Cator}\ and\ \citenamefont
  {Van~Mieghem}(2012)}]{PhysRevE.85.056111}%
  \BibitemOpen
  \bibfield  {author} {\bibinfo {author} {\bibfnamefont {E.}~\bibnamefont
  {Cator}}\ and\ \bibinfo {author} {\bibfnamefont {P.}~\bibnamefont
  {Van~Mieghem}},\ }\bibfield  {title} {\bibinfo {title} {Second-order
  mean-field susceptible-infected-susceptible epidemic threshold},\ }\href
  {https://doi.org/10.1103/PhysRevE.85.056111} {\bibfield  {journal} {\bibinfo
  {journal} {Phys. Rev. E}\ }\textbf {\bibinfo {volume} {85}},\ \bibinfo
  {pages} {056111} (\bibinfo {year} {2012})}\BibitemShut {NoStop}%
\bibitem [{\citenamefont {Gillespie}(1977)}]{Gillespie}%
  \BibitemOpen
  \bibfield  {author} {\bibinfo {author} {\bibfnamefont {D.~T.}\ \bibnamefont
  {Gillespie}},\ }\bibfield  {title} {\bibinfo {title} {Exact stochastic
  simulation of coupled chemical reactions},\ }\href
  {https://doi.org/10.1021/j100540a008} {\bibfield  {journal} {\bibinfo
  {journal} {J. Phys. Chem.}\ }\textbf {\bibinfo {volume} {81}},\ \bibinfo
  {pages} {2340} (\bibinfo {year} {1977})}\BibitemShut {NoStop}%
\end{thebibliography}%

\end{document}